\documentclass[twocolumn]{aastex631}

\usepackage{float}
\usepackage{amsmath} 
\usepackage{relsize} 
\usepackage{multirow}
\newcommand{\sbsc}[1]{_\mathrm{#1}}

\accepted{December 7, 2023}
\submitjournal{ApJS}

\shorttitle{The PHANGS-AstroSat Atlas}
\shortauthors{Hassani et al.}
\graphicspath{{./}{figures/}}
\usepackage{xspace}

\begin{document}

\title{The PHANGS-AstroSat Atlas of Nearby Star Forming Galaxies}

\correspondingauthor{Hamid Hassani}
\email{hhassani@ualberta.ca}

\author[0000-0002-8806-6308]{Hamid Hassani}
\affiliation{Dept. of Physics, University of Alberta, 4-183 CCIS, Edmonton, Alberta, T6G 2E1, Canada}

\author[0000-0002-5204-2259]{Erik Rosolowsky}
\affiliation{Dept. of Physics, University of Alberta, 4-183 CCIS, Edmonton, Alberta, T6G 2E1, Canada}

\author[0000-0001-9605-780X]{Eric~W.~Koch}
\affiliation{Center for Astrophysics $\mid$ Harvard \& Smithsonian, 60 Garden St., 02138 Cambridge, MA, USA}
\affiliation{Dept. of Physics, University of Alberta, 4-183 CCIS, Edmonton, Alberta, T6G 2E1, Canada}

\author[0000-0002-3025-1412]{Joseph Postma}
\affiliation{Dept. of Physics and Astronomy, University of Calgary, Calgary, Alberta, Canada}

\author[0000-0002-9553-953X]{Joseph Nofech}
\affiliation{Dept. of Physics, University of Alberta, 4-183 CCIS, Edmonton, Alberta, T6G 2E1, Canada}

\author[0009-0007-7949-6633]{Harrisen Corbould}
\affiliation{Dept. of Physics, University of Alberta, 4-183 CCIS, Edmonton, Alberta, T6G 2E1, Canada}

\author[0000-0002-8528-7340]{David Thilker}
\affiliation{Center for Astrophysical Sciences, The Johns Hopkins University, Baltimore, MD USA}

\author[0000-0002-2545-1700]{Adam~K.~Leroy}
\affiliation{Department of Astronomy, The Ohio State University, Columbus, Ohio 43210, USA}

\author[0000-0002-3933-7677]{Eva Schinnerer}
\affiliation{Max Planck Institut für Astronomie, Königstuhl 17, D-69117 Heidelberg, Germany}

\author[0000-0002-2545-5752]{Francesco~Belfiore}
\affiliation{
INAF — Osservatorio Astrofisico di Arcetri, Largo E. Fermi 5, I-50125, Florence, Italy}

\author[0000-0003-0166-9745]{Frank Bigiel}
\affiliation{Argelander-Institut f\"ur Astronomie, Universit\"at Bonn, Auf dem H\"ugel 71, 53121 Bonn, Germany}

\author[0000-0003-0946-6176]{Médéric Boquien}
\affiliation{Instituto de Alta Investigación, Universidad de Tarapacá, Casilla 7D, Arica, Chile}

\author[0000-0002-5635-5180]{M\'elanie Chevance}
\affiliation{Zentrum f\"{u}r Astronomie der Universit\"{a}t Heidelberg, Institut f\"{u}r Theoretische Astrophysik, Albert-Ueberle-Str. 2, 69120 Heidelberg}
\affiliation{Cosmic Origins Of Life (COOL) Research DAO, coolresearch.io}

\author[0000-0002-5782-9093]{Daniel~A.~Dale}
\affiliation{Department of Physics \& Astronomy, University of Wyoming, Laramie, WY 82071 USA}

\author[0000-0002-4755-118X]{Oleg V. Egorov}
\affiliation{Astronomisches Rechen-Institut, Zentrum f\"{u}r Astronomie der Universit\"{a}t Heidelberg, M\"{o}nchhofstr. 12-14, D-69120 Heidelberg, Germany}

\author[0000-0002-6155-7166]{Eric Emsellem}
\affiliation{European Southern Observatory, Karl-Schwarzschild-Stra{\ss}e 2, 85748 Garching, Germany}
\affiliation{Univ Lyon, Univ Lyon1, ENS de Lyon, CNRS, Centre de Recherche Astrophysique de Lyon UMR5574, F-69230 Saint-Genis-Laval France}

\author[0000-0001-6708-1317]{Simon C.~O.~Glover}
 \affiliation{Universit\"{a}t Heidelberg, Zentrum f\"{u}r Astronomie, Institut f\"{u}r Theoretische Astrophysik, Albert-Ueberle-Str. 2, 69120, Heidelberg, Germany}
 
\author[0000-0002-3247-5321]{Kathryn~Grasha}
\altaffiliation{ARC DECRA Fellow}
\affiliation{
Research School of Astronomy and Astrophysics, Australian National University, Canberra, ACT 2611, Australia}
\affiliation{ARC Centre of Excellence for All Sky Astrophysics in 3 Dimensions (ASTRO 3D), Australia}

\author[0000-0002-9768-0246]{Brent Groves}
\affiliation{International Centre for Radio Astronomy Research, University of Western Australia, 7 Fairway, Crawley, 6009 WA, Australia}

\author[0000-0001-7448-1749]{Kiana Henny}
\affiliation{Department of Physics \& Astronomy, University of Wyoming, Laramie, WY 82071 USA}

\author[0000-0002-0432-6847]{Jaeyeon Kim}
\affiliation{Kavli Institute for Particle Astrophysics \& Cosmology (KIPAC), Stanford University, CA 94305, USA} 

\author[0000-0002-0560-3172]{Ralf S.\ Klessen}
 \affiliation{Universit\"{a}t Heidelberg, Zentrum f\"{u}r Astronomie, Institut f\"{u}r Theoretische Astrophysik, Albert-Ueberle-Str. 2, 69120, Heidelberg, Germany.}\affiliation{Universit\"{a}t Heidelberg, Interdisziplin\"{a}res Zentrum f\"{u}r Wissenschaftliches Rechnen, Im Neuenheimer Feld 205, D-69120 Heidelberg, Germany.}

\author[0000-0001-6551-3091]{Kathryn~Kreckel}
\affiliation{Astronomisches Rechen-Institut, Zentrum f\"{u}r Astronomie der Universit\"{a}t Heidelberg, M\"{o}nchhofstra\ss e 12-14, 69120 Heidelberg, Germany}

\author[0000-0002-8804-0212]{J.~M.~Diederik Kruijssen}
\affiliation{Technical University of Munich, School of Engineering and Design, Department of Aerospace and Geodesy, Chair of Remote Sensing Technology, Arcisstr. 21, 80333 Munich, Germany}
\affiliation{Cosmic Origins Of Life (COOL) Research DAO, coolresearch.io}

\author[0000-0002-2278-9407]{Janice C. Lee}
\affiliation{Space Telescope Science Institute, 3700 San Martin Drive, Baltimore, MD 21218, USA}
\affiliation{Steward Observatory, University of Arizona, Tucson, AZ 85712, USA}

\author[0000-0002-1790-3148]{Laura A. Lopez}
\affiliation{Department of Astronomy, The Ohio State University, 140 W. 18th Ave., Columbus, OH 43210, USA}
\affiliation{Center for Cosmology and AstroParticle Physics, The Ohio State University, 191 W. Woodruff Ave., Columbus, OH 43210, USA}

\author[0000-0002-3289-8914]{Justus Neumann}
\affiliation{Max-Planck-Institut f\"{u}r Astronomie, K\"{o}nigstuhl 17, D-69117 Heidelberg, Germany}

\author[0000-0002-1370-6964]{Hsi-An Pan}
\affiliation{Department of Physics, Tamkang University, No.151, Yingzhuan Road, Tamsui District, New Taipei City 251301, Taiwan}

\author[0000-0002-4378-8534]{Karin M. Sandstrom}
\affiliation{Department of Astronomy \& Astrophysics, University of California, San Diego, 9500 Gilman Drive, La Jolla, CA 92093}

\author[0000-0002-4781-7291]{Sumit K. Sarbadhicary}
\affiliation{Department of Physics, The Ohio State University, Columbus, Ohio 43210, USA}
\affiliation{Center for Cosmology \& Astro-Particle Physics, The Ohio State University, Columbus, Ohio 43210, USA}

\author[0000-0003-0378-4667]{Jiayi~Sun}
\affiliation{Department of Physics and Astronomy, McMaster University, 1280 Main Street West, Hamilton, ON L8S 4M1, Canada}
\affiliation{Canadian Institute for Theoretical Astrophysics (CITA), University of Toronto, 60 St George Street, Toronto, ON M5S 3H8, Canada}

\author[0000-0002-0012-2142]{Thomas G. Williams}
\affiliation{Sub-department of Astrophysics, Department of Physics, University of Oxford, Keble Road, Oxford OX1 3RH, UK}




\begin{abstract}

We present the Physics at High Angular resolution in Nearby GalaxieS (PHANGS)-{\it AstroSat} atlas, which contains ultraviolet imaging of 31 nearby star-forming galaxies captured by the Ultraviolet Imaging Telescope (UVIT) on the  \textit{AstroSat} satellite. The atlas provides a homogeneous data set of far- and near-ultraviolet maps of galaxies within a distance of 22 Mpc and a median angular resolution of 1.4 arcseconds (corresponding to a physical scale between 25 and 160 pc). After subtracting a uniform ultraviolet background and accounting for Milky Way extinction, we compare our estimated flux densities to GALEX observations, finding good agreement. We find candidate extended UV disks around the galaxies NGC 6744 and IC 5332. We present the first statistical measurements of the clumping of the UV emission and compare it to the clumping of molecular gas traced with ALMA. We find that bars and spiral arms exhibit the highest degree of clumping, and the molecular gas is even more clumped than the FUV emission in galaxies. We investigate the variation of the ratio of observed FUV to H$\alpha$ in different galactic environments and kpc-sized apertures. We report that $\sim 65$\% variation of the $\log_{10}$(FUV/H$\alpha$) can be described through a combination of dust attenuation with star formation history parameters. The PHANGS-{\it AstroSat} atlas enhances the multi-wavelength coverage of our sample, offering a detailed perspective on star formation. When integrated with PHANGS data sets  from ALMA, VLT-MUSE, HST and JWST, it develops our comprehensive understanding of attenuation curves and dust attenuation in star-forming galaxies.\end{abstract}



\section{Introduction} \label{sec:intro}
The high-mass stellar population of a galaxy plays a central role in galactic evolution. High-mass stars, which are short-lived and must have formed recently, change the colors of galaxies and provide the mechanical and radiative feedback that determine the future evolution of the system \citep{krumholz2014}.  Beyond the Local Group, our study of these young, massive stars relies on their photospheric emission, which is directly traced in the near- and far-ultraviolet and indirectly observed using recombination line emission in the optical (e.g., H$\alpha$) and reprocessing from dust \citep{kennicutt12}.

Studying ultraviolet (UV) emission from nearby extragalactic targets traces star formation over a longer time scale (100-200 Myr) in comparison to other typical tracers, such as recombination lines \citep[5-20 Myr;][]{kennicutt12}. To understand the UV spectral behaviour of young stellar populations located beyond the optical disk, both large fields-of-view (FoV) and multi-band UV imaging are essential. The latter feature is a key to constraining the attenuation properties along with the UV color \citep{Decleir}. Moreover, because of its sensitivity to small numbers of high-mass stars, UV emission provides a good tracer of star formation beyond optical disks \citep{Thilker2007}, and more generally in galactic environments characterized by low density \citep{lee11} such as dwarf galaxies. Studying short-lived massive stars in low-density outskirts of galaxies constrains how gas may accrete into galaxies and which mechanism is responsible for the regulation of star formation in these low-column density environments \citep{GildePaz2007}.


In the past decades, several telescopes such as the Galaxy Evolution Explorer (GALEX) and the Neil Gehrels Swift Observatory's Ultraviolet Optical Telescope (UVOT) have mapped UV bands in nearby galaxies \citep{lee11,swift, z0mgs}. GALEX provided imaging in the near-UV (NUV; 2500 \AA) and far-UV (FUV; 1500 \AA) with a field of view of $\sim$1.28$^{\circ}$ and angular resolution of 4 to 6\arcsec, respectively 
\citep{Martin2005}. On the other hand, Swift-UVOT has three UV bands with a smaller FoV (17\arcmin) and an angular resolution of $\sim$2.5\arcsec \citep{Roming}. Many insights about nearby galaxy populations have been derived from GALEX, combined with multiwaveband results, which have provided a rich understanding of galaxy evolution and the impacts of star formation \citep[e.g.,][]{Salim2007,GildePaz2007_GALEX_ATLAS, Leroy_2008, lee2009}.

The Ultraviolet Imaging Telescope (UVIT) on the {\it AstroSat} mission offers an improvement in  resolution for imaging energetic UV photons with multiple UV bands. It extends the legacy of GALEX/SWIFT by effectively resolving the structures of star-forming regions, thanks to its $<1.8\arcsec$\,resolution \citep{uvit,astrosat_MISSION}. UVIT observes in 9 bands, covering the range from 1480\,\AA\ to 2790\,\AA, with a 28\arcmin\ FoV  \citep{astrosat-performance-Rahna}. UVIT data provide a new view of the resolved properties of faint stellar populations beyond the optical disk of galaxies and offer detailed additional information on the UV colour and properties of the underlying stellar population.


The UV filter set of the {\it AstroSat} UVIT instrument consists of one narrow-band filter (N279N), three wide-band filters (F148W, F154W, and N242W), and five medium-band filters (F169M, F172M, N219M, N245M, and N263M), where filter names follow the {\it Hubble} Space Telescope convention that the three digit number represents the central wavelength in nm. These filters capture photospheric emission from stellar populations with transmission curves in Figure \ref{fig:filterset} \citep{Leahy_2022,Leahy_2022_1}. The data from AstroSat UVIT can be utilized to discern and model both the age and mass of young stellar populations, ranging from approximately $5-200$ Myr and with masses greater than $10^{3-4} M_{\odot}$  \citep[e.g.;][]{Ujjwal}. Many of these populations are located beyond $R_{25}$, i.e.\ the radius at which the surface brightness falls below 25 magnitudes per square arcsecond in the B-band.



Sensitive, wide-field UV observations are even more powerful when combined with probes of the different phases of the ISM and stellar populations. The star formation process, as traced by the NUV and FUV, offers valuable insight when aligned with other tracers of star formation phases. CO observations, which trace molecular gas, identify dust-shrouded star-forming regions, areas where UV photons can quickly destroy clumps. Observations of ionized gas, captured through hydrogen recombination lines, showcase the re-emission stemming from the Extreme UV (EUV) of the highest mass stars.

Integrating the UV observations from nearby galaxies with a spatial resolution below 200 pc can enhance the comprehensive understanding advanced by projects like the Physics at High Angular resolution in Nearby GalaxieS (PHANGS) collaboration.  The PHANGS-ALMA survey \citep{phangsalma} yielded $\sim1\arcsec$ resolution observations of CO(2-1) emission from molecular clouds in 90 nearby galaxies.  Similarly, PHANGS-MUSE used VLT/MUSE observations to map out optical emission line from 19 of the 90 ALMA targets, also at $\lesssim 1\arcsec$ resolution.  The PHANGS-{\it AstroSat} provides an essential complement to the UV imaging obtained by the PHANGS-HST survey \citep{phangs-hst} which only probes to $\sim$2500\AA, but achieves an angular resolution of 0\farcs08 over the smaller fields-of-view of the WFC3 camera. In this work, we compare the new UV data to these other tracers of the star formation process to understand how galaxies evolve at these arcsecond scales.


One of the main drivers of galaxy evolution is the relationship between gas and star formation called the Kennicutt-Schmidt (KS) relation \citep{Kennicutt98}, which presents a tight correlation between the surface densities of gas and star formation on large scales ($>1~\mathrm{kpc}$) in galaxies. However, several studies indicate that this relation operates on kpc scales and it may break down at smaller scales \cite[e.g.][]{Bigiel2008,Kruijssen2019}. While these results typically rely on recombination line tracers of star formation, {\it AstroSat} imaging can play a key role in estimating the link between star formation and the amount of molecular gas over longer timescales.  UV observation can make progress even in the outskirts of galaxies, where the fraction of molecular relative to neutral gas is low. However, we should note that simulations such as those by \cite{Khoperskov} have shown that the relation between UV-based star formation rates and the total amount of gas at 50-pc resolution is unclear, and observations suggest that the KS relation prescription cannot be extrapolated to the low-density environments of outer galaxy disks \citep{Dessauges}. These findings underscore the need for high-resolution investigations of UV-based star formation processes beyond the optical disks of galaxies. 

The relationship between UV and H$\alpha$ luminosities provides valuable information on dust content, the age of stellar populations, and the metallicity of galaxies. Previous studies \citep[e.g., ][]{lee2009, lee11} have attempted to uncover the details of the UV-H$\alpha$ relation with a focus on understanding recipes for the integrated star formation rate in galaxies. However, the dependence of this relation on the actual amount of molecular gas and dust reddening remains unclear. Furthermore, there is a lack of detailed information on how the ratio changes across different environments within galaxies, such as bars, centres, and spiral arms at small scales.  The ionized gas traced by H$\alpha$ and observed through MUSE/VLT observations with the PHANGS-MUSE survey offers a new perspective on recent star formation up to 10 million years ago as well as dust reddening \citep{phangs-muse}.  When combined with the new UVIT observations, we will be able to directly relate changing UV-to-H$\alpha$ to the local environment in these galaxies.


In this paper, we present a homogeneous data set of UV images of 31 nearby galaxies observed by {\it AstroSat} in the wavelength range of 1480\,\AA\, to 2790\,\AA. Our main focus is to reduce and assess the quality of the data set and compare it with previous low-resolution observations. Additionally, we study the variation of the UV and molecular gas clumpiness in different morphological regions of galaxies. By comparing the FUV to H$\alpha$ flux in different galaxies, we aim to provide predictive models for the ratio in different environments. This work is expected to contribute to the interpretation of different tracers of star formation at high resolution ($<$200 pc) and provide a quantitative framework for the relation between UV and H$\alpha$ luminosity.
 
In Section \ref{sec:sample_selection}, we present our sample selection strategies and then the observations. We focus on the reduction and quality assessment of our data in Section \ref{sec:procs}. Then, we present two new XUV disk candidates in Section \ref{sec:xuv} and provide details about the clumping of UV and CO emission at high resolution ($\lesssim 1.5''$) in Section \ref{sec:clumping}. Finally, we compare the luminosity of FUV to H$\alpha$ emission to show how these two star formation tracers vary in different morphological parts of galaxies and how their ratio depends on dust attenuation, star formation history parameters, and the equivalent width (EW) of the H$\alpha$ emission (Section \ref{sec:fuvha}).

\section{Data}
\label{sec:data}
\subsection{Sample Selection}
\label{sec:sample_selection}
To maximize the opportunity for multiwavelength science, we selected galaxies that have been observed as part of the PHANGS survey for observation using {\it AstroSat}.  The primary PHANGS sample consists of 90 nearby galaxies observed with the Atacama Large Millimeter/submillimeter Array \citep[ALMA,][]{Leroy2021} in molecular line emission.  These targets are selected to be nearby ($D<22~\mathrm{Mpc}$) so a high efficiency observation with ALMA yields high quality maps of the molecular interstellar medium with $\sim 1''$ resolution. The proximity of the targets means that the $1''$ resolution translates into small physical scales, providing a robust census of star-forming galaxies at $\sim 100$ pc linear resolution, which matches the typical scale heights of galaxies \citep{sun20} and the size of giant molecular clouds \citep{rosolowsky21}. The PHANGS targets are relatively massive with ($10^{9.75} < M_\star / M_\odot < 10^{11.0}$,) and mostly lie on the star forming main sequence, with SFR/$M_\star > 10^{-11}~\mathrm{yr}^{-1}$. The sample was selected with a constraint to avoid highly inclined systems, ensuring that $i< 75^\circ$.  Table \ref{tab:galpropss} presents the 31 galaxies in the PHANGS-{\it AstroSat} sample and their properties, adopted from \cite{Leroy2021}.

\subsection{Observations}
\label{sec:observations}
The {\it AstroSat} satellite mission provides observations from four instruments: the Large Area X-ray Proportional Counter (LAXPC), the Cadmium Zinc Telluride Imager (CZTI), the Soft X-ray Telescope (SXT), and the Ultraviolet Imaging Telescope \citep[UVIT;][]{uvit}, which is the primary instrument used in this study. The main goal of UVIT is imaging in far-ultraviolet (FUV; 1300$ - $1800\,\AA) and near-ultraviolet (NUV; 2000$ - $3000\,\AA) with a field of view of $\sim$28$\arcmin$. We observed 11 of the 31 selected galaxies under programs T03-032, T03-033, A05-022, A07-027, A08-003, and A10-021.
The PHANGS-{\it AstroSat} observations adopted a uniform strategy for the primary sample, requesting 3.6 ks of time on each target in the F148W filter.  However, instrumental fluctuations and orbital accessibility limit the actual time acquired to \(1-3 \) ks.  We supplement these observations with additional archival data for 20 additional PHANGS targets, identified from the {\it AstroSat} Archive\footnote{{\url{https://astrobrowse.issdc.gov.in/astro_archive/archive/Home.jsp}}}. As the supplementary data come from various observing programs, they use different observational strategies and exposures.  Archival observations feature a variety of adopted filters and exposure times, some of which may be longer than those proposed for our observations. For instance, NGC 4476 has an exposure time that is 10 times longer than most of the sample.
During UVIT observations, {\it AstroSat} consistently records data from other instruments within its focal plane. While we do not present these data in this work, they are accessible directly from the archive.

\begin{figure}
    \centering
    \includegraphics[width=0.95\columnwidth]{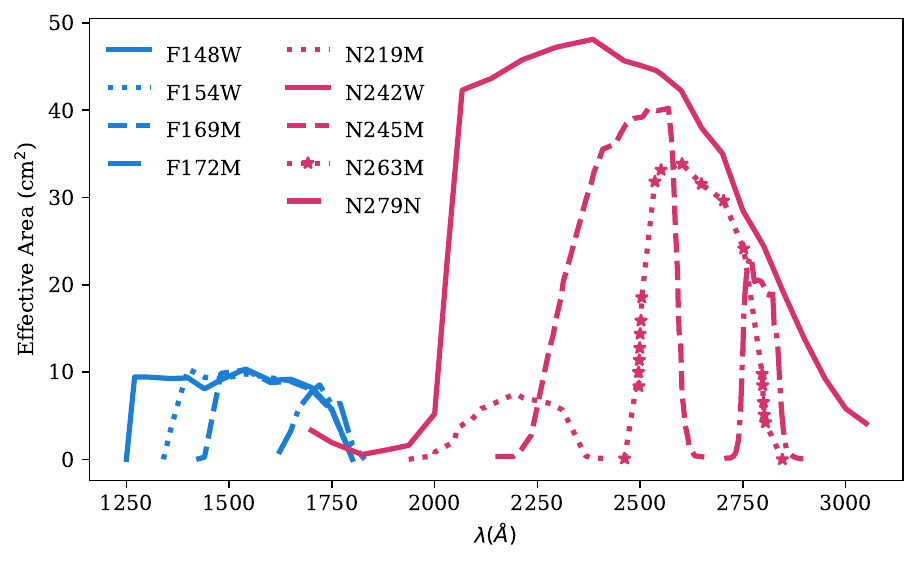}
    \caption{Filters used in the {\it AstroSat} UVIT observations, corrected for in-orbit calibrations. Blue lines show the effective area of FUV bands, whereas the pink lines are the NUV bands.}
    \label{fig:filterset}
\end{figure}

The combination of archival and dedicated observations lead to heterogeneous data across the 31 galaxies.  Table \ref{tab:summary} summarizes the available {\it AstroSat} observations in this atlas, including the observation IDs, dates of observations, and the observed filter set as illustrated in Figure \ref{fig:filterset}.  This table includes allocated projects led by our team, as well as data sets retrieved from the archive.  Table \ref{tab:exposures} summarizes the exposure times for the different galaxies across the different filter sets. We describe the data processing for UVIT in Section \ref{sec:procs}.

\begin{deluxetable*}{ccccl}
\setlength{\tabcolsep}{2pt} 
\tablecaption{PHANGS-{\it AstroSat} Galaxies\label{tab:summary}}
\tablewidth{0pt}
\tablehead{
\colhead{Galaxy} & \colhead{Observation ID} & 
 \colhead{PI$^{*}$} & \colhead{Observation Date} & \colhead{Filter(s)} 
}
\startdata
IC5332 & A07\_027T02\_9000003258 & rosolowsky  & 26-Oct-2019 & F148W \\
 & A07\_027T02\_9000003640 & rosolowsky & 4-May-2020 & F148W \\
NGC0253 & G0685\_010T01\_9000001672 & jmurthy & 08-Nov-2017 & F169M,N245M \\
 & G0685\_031T01\_9000001702 & askpati & 19-Nov-2017 & F169M,N219M,N263M \\
NGC0300 & G05\_235T01\_9000000590 & hutchingsj & 11-Aug-2016 & F148W,F154W,F169M,F172M,N219M,N245M,N263M \\
NGC0628 & A04\_209T01\_9000002378 & carobert & 20-Sep-2018 & F154W,F172M \\
 & G06\_151T01\_9000000836 & askpati &  29-Nov-2016 & F148W,F154W,F169M,F172M,N242W,N219M,N245M,N263M,N279N \\
NGC1097 & A10\_021T01\_9000004044 & rosolowsky & 02-Dec-2020 & F148W \\
NGC1300 & A07\_027T05\_9000003506 & rosolowsky & 16-Feb-2020 & F148W \\
NGC1317 & A04\_164T01\_9000001770 & nilkanth & 14-Dec-2017 & F148W,F154W,F169M,F172M,N219M \\
NGC1365 & A02\_006T01\_9000000776 & gulabd & 08-Nov-2016 & F169M,N279N \\
 & A02\_006T01\_9000000802 & gulabd & 17-Nov-2016 & F148W,F169M,N279N \\
 & A02\_006T01\_9000000934 & gulabd & 28-Dec-2016 & F169M,N279N \\
 & G07\_057T02\_9000001504 & stalin & 31-Aug-2017 & F148W,F172M,N219M,N263M \\
 NGC1385 & A07\_027T06\_9000003508 & rosolowsky & 16-Feb-2020 & F148W \\
NGC1433 & G07\_066T01\_9000001510 & swarna & 01-Sep-2017 & F154W,F169M,N219M,N245M,N263M,N279N \\
NGC1512 & G06\_135T01\_9000000908 & swarna & 21-Dec-2016 & F154W,N245M,N263M \\
 & G07\_068T01\_9000001502 & kanak & 30-Aug-2017 & F154W,N242W \\
NGC1546 & A07\_010T11\_9000003240 & rrampazzo & 17-Oct-2019 & F148W \\
NGC1566 & G06\_087T01\_9000000926 & stalin & 26-Dec-2016 & F148W,F172M,N219M,N263M \\
 & T02\_085T01\_9000002296 & gulabd & 11-Aug-2018 & F154W \\
 & T03\_020T01\_9000002444 & gulabd & 22-Oct-2018 & F154W \\
NGC2090 & A05\_155T02\_9000003200 & mousumi & 24-Sep-2019 & F148W \\
NGC2835 & T03\_032T01\_9000002564 & rosolowsky & 14-Dec-2018 & F148W \\
NGC2903 & G08\_031T03\_9000001972 & askpati & 12-Mar-2018 & F148W,F169M,N219M,N263M \\
NGC3351 & T03\_034T01\_9000002500 & rosolowsky & 10-Nov-2018 & F148W \\
NGC3621 & G08\_083T03\_9000002022 & stalin & 07-Apr-2018 & F148W,F172M \\
NGC3627 & T03\_033T01\_9000002568 & rosolowsky & 15-Dec-2018 & F148W \\
NGC4254 & A08\_003T04\_9000003634 & hutchingsj & 02-May-2020 & F148W \\
NGC4298 & A10\_021T04\_9000004116 & rosolowsky & 17-Jan-2021 & F148W \\
NGC4321 & A08\_003T05\_9000003426 & hutchingsj & 11-Jan-2020 & F154W \\
NGC4476 & G06\_051T01\_9000000972 & pcote\_nrc & 22-Jan-2017 & F154W,N242W \\
NGC4535 & A10\_021T06\_9000004338 & rosolowsky	 & 27-April-2021 & F148W \\
NGC4571 & G06\_016T01\_9000001052 & kanak & 25-Feb-2017 & F154W,N263M \\
NGC4579 & A08\_003T09\_9000003644 & hutchingsj & 05-May-2020 & F154W \\
NGC4654 & A07\_027T12\_9000003664 & rosolowsky & 13-May-2020 & F148W \\
 & A08\_003T08\_9000003638 & hutchingsj & 03-May-2020 & F154W \\
NGC5128 & G08\_023T01\_9000001978 & sreekumar & 15-Mar-2018 & F148W,N219M,N245M,N279N \\
NGC6744 & A05\_022T09\_9000003058 & rosolowsky & 24-July-2019 & F148W \\
 & A10\_021T10\_9000004210 & rosolowsky & 26-Feb-2021 & F148W \\
NGC7496 & A07\_027T16\_9000003222 & rosolowsky & 05-Oct-2019 & F148W \\
 & A07\_027T16\_9000003642 & rosolowsky	 &  05-May-2020 & F148W \\
 & A07\_027T16\_9000003666  & rosolowsky & 14-May-2020 & F148W \\
NGC7793 & G06\_024T01\_9000000780 & annapurni & 10-Nov-2016 & F148W,N242W
\enddata
\tablecomments{$^{*}$ The PIs listed here are directly read from the {\it AstroSat} Archive Search system. The `Filter(s)' column shows all available filters as presented on the {\it AstroSat} Archive Search; we do not necessarily utilize calibrated data for every single listed filter for each galaxy.}
\end{deluxetable*}
\begin{deluxetable*}{crrrrrrrrr}
\tablecaption{PHANGS-{\it AstroSat} Exposures\label{tab:exposures}}
\tablewidth{0pt}
\tablehead{
\colhead{Galaxy} &
\colhead{F148W	} &
\colhead{F154W	} &
\colhead{F172M	} &
\colhead{F169M	} &
\colhead{N242W	} &
\colhead{N245M	} &
\colhead{N263M	} &
\colhead{N219M} &\colhead{N279N } \\
\colhead{} & \colhead{(ks)} & \colhead{(ks)} & \colhead{(ks)} & \colhead{(ks)} & \colhead{(ks)} & \colhead{(ks)} & \colhead{(ks)} & \colhead{(ks)} & \colhead{(ks)}
}
\startdata
IC5332 & 2.28$^{*}$ & \nodata & \nodata & \nodata & \nodata & \nodata & \nodata & \nodata & \nodata \\
NGC0253 & \nodata & \nodata & \nodata & 12.29 & \nodata & 12.40 & 3.02 & 4.57 & \nodata \\
NGC0300 & 12.84 & 1.65 & 0.95 & 6.92 & \nodata & 2.74 & 2.52 & 0.32 & \nodata \\
NGC0628 & 1.82 & 4.88 & 19.12 & 2.42 & 0.43 & 1.32 & 2.08 & 1.36 & 4.65 \\
NGC1097 & 1.75$^{*}$ & \nodata & \nodata & \nodata & \nodata & \nodata & \nodata & \nodata & \nodata \\
NGC1300 & 3.44$^{*}$ & \nodata & \nodata & \nodata & \nodata & \nodata & \nodata & \nodata & \nodata \\
NGC1317 & 0.83 & 1.01 & 0.87 & 1.41 & \nodata & \nodata & \nodata & 4.32 & \nodata \\
NGC1365 & 2.10 & \nodata & 6.59 & 46.03 & \nodata & \nodata & 2.07 & 6.53 & 47.39 \\
NGC1385 & 1.79$^{*}$  & \nodata & \nodata  & \nodata & \nodata & \nodata & \nodata & \nodata  & \nodata  \\
NGC1433 & \nodata & 3.14 & \nodata & 3.17 & \nodata & 1.67 & 0.91 & 3.20 & 1.25 \\
NGC1512 & \nodata & 5.80 & \nodata & \nodata & 3.70 & 1.02 & 1.20 & \nodata & \nodata \\
NGC1546 & 6.65 & \nodata & \nodata & \nodata & \nodata & \nodata & \nodata & \nodata & \nodata \\
NGC1566 & 2.94 & 3.38 & 1.34 & \nodata & \nodata & \nodata & 2.96 & 1.37 & \nodata \\
NGC2090 & 6.43 & \nodata & \nodata & \nodata & \nodata & \nodata & \nodata & \nodata & \nodata \\
NGC2835 & 3.41$^{*}$ & \nodata & \nodata & \nodata & \nodata & \nodata & \nodata & \nodata & \nodata \\
NGC2903 & 3.50 & \nodata & \nodata & 4.32 & \nodata & \nodata & 3.35 & 4.61 & \nodata \\
NGC3351 & 2.64$^{*}$ & \nodata & \nodata & \nodata & \nodata & \nodata & \nodata & \nodata & \nodata \\
NGC3621 & 2.10 & \nodata & 6.37 & \nodata & \nodata & \nodata & \nodata & \nodata & \nodata \\
NGC3627 & 3.17$^{*}$ & \nodata & \nodata & \nodata & \nodata & \nodata & \nodata & \nodata & \nodata \\
NGC4254 & \nodata & 7.60 & \nodata & \nodata & \nodata & \nodata & \nodata & \nodata & \nodata \\
NGC4298 & 2.92$^{*}$ & \nodata & \nodata & \nodata & \nodata & \nodata & \nodata & \nodata & \nodata \\
NGC4321 & \nodata & 7.63 & \nodata & \nodata & \nodata & \nodata & \nodata & \nodata & \nodata \\
NGC4476 & \nodata & 35.88 & \nodata & \nodata & 35.21 & \nodata & \nodata & \nodata & \nodata \\
NGC4535 & 1.21$^{*}$  & \nodata & \nodata  & \nodata & \nodata & \nodata & \nodata & \nodata  & \nodata  \\
NGC4571 & \nodata & 9.58 & \nodata & \nodata & \nodata & \nodata & 9.65 & \nodata & \nodata \\
NGC4579 & \nodata & 7.59 & \nodata & \nodata & \nodata & \nodata & \nodata & \nodata & \nodata \\
NGC4654 & 3.52$^{*}$ & \nodata & \nodata & \nodata & \nodata & \nodata & \nodata & \nodata & \nodata \\
NGC5128 & 23.31 & \nodata & \nodata & \nodata & \nodata & 4.75 & \nodata & 10.53 & 8.00 \\
NGC6744 & 3.45$^{*}$ & \nodata & \nodata & \nodata & \nodata & \nodata & \nodata & \nodata & \nodata \\
NGC7496 & 6.16$^{*}$ & \nodata & \nodata & \nodata & \nodata & \nodata & \nodata & \nodata & \nodata \\
NGC7793 & 7.53 & \nodata & \nodata & \nodata & 8.10 & \nodata & \nodata & \nodata & \nodata
\enddata
\tablecomments{$^{*}$ Observed as a part of the PHANGS-{\it AstroSat} request.}
\end{deluxetable*}

\begin{deluxetable}{cccccc}
\tablecaption{Adopted Galaxy Properties}
\label{tab:galpropss}
\tablehead{
\colhead{Galaxy} & \colhead{Stellar Mass} & \colhead{SFR} & \colhead{Distance} & Incl. & PA \\
\colhead{} & \colhead{($10^{10} M_\odot$)}  & \colhead{(M$_{\odot}$ yr$^{-1}$)} &  \colhead{(Mpc)} & \colhead{($^\circ$)} & \colhead{($^\circ$)}
}
\startdata
IC5332 & 0.47 & 0.41 & 9.01 & 26.9 & 74.4 \\
NGC0253 & 4.34 & 5.0 & 3.7 & 75.0 & 52.48 \\
NGC0300 & 0.18 & 0.15 & 2.09 & 39.8 & 114.3 \\
NGC0628 & 2.19 & 1.75 & 9.84 & 8.9 & 20.7 \\
NGC1097$^{*}$ & 5.75 & 4.74 & 13.58 & 48.6 & 122.4 \\
NGC1300 & 4.14 & 1.17 & 18.99 & 31.8 & 278.0 \\
NGC1317 & 4.17 & 0.48 & 19.11 & 23.2 & 221.5 \\
NGC1365$^{*}$ & 9.78 & 16.9 & 19.57 & 55.4 & 201.1 \\
NGC1385 & 0.95 & 2.09 & 17.22 & 44.0 & 181.3 \\
NGC1433 & 7.34 & 1.13 & 18.63 & 28.6 & 199.7 \\
NGC1512 & 5.16 & 1.28 & 18.83 & 42.5 & 261.9 \\
NGC1546 & 2.24 & 0.83 & 17.69 & 70.3 & 147.8 \\
NGC1566$^{*}$ & 6.09 & 4.54 & 17.69 & 29.5 & 214.7 \\
NGC2090 & 1.09 & 0.41 & 11.75 & 64.5 & 192.46 \\
NGC2835 & 1.0 & 1.24 & 12.22 & 41.3 & 1.0 \\
NGC2903 & 4.3 & 3.08 & 10.0 & 66.8 & 203.7 \\
NGC3351 & 2.3 & 1.32 & 9.96 & 45.1 & 193.2 \\
NGC3621$^{*}$ & 1.14 & 0.99 & 7.06 & 65.8 & 343.8 \\
NGC3627$^{*}$ & 6.81 & 3.84 & 11.32 & 57.3 & 173.1 \\
NGC4254 & 2.66 & 3.07 & 13.1 & 34.4 & 68.1 \\
NGC4298 & 1.05 & 0.46 & 14.92 & 59.2 & 313.9 \\
NGC4321 & 5.56 & 3.56 & 15.21 & 38.5 & 156.2 \\
NGC4476 & 0.65 & 0.04 & 17.54 & 60.14 & 27.38 \\
NGC4535 & 3.4 & 2.16 & 15.77 & 44.7 & 179.7 \\
NGC4571 & 1.23 & 0.29 & 14.9 & 32.7 & 217.5 \\
NGC4579$^{*}$ & 13.99 & 2.17 & 21.0 & 40.22 & 91.3 \\
NGC4654 & 3.69 & 3.79 & 21.98 & 55.6 & 123.2 \\
NGC5128 & 9.38 & 1.23 & 3.69 & 45.33 & 32.17 \\
NGC6744 & 5.29 & 2.41 & 9.39 & 52.7 & 14.0 \\
NGC7496$^{*}$ & 0.99 & 2.26 & 18.72 & 35.9 & 193.7 \\
NGC7793 & 0.23 & 0.27 & 3.62 & 50.0 & 290.0 \\
\enddata
\tablecomments{$^{*}$ host an AGN based on \citet{agn_galaxies}. Properties are taken from \citet{Leroy2021}.}
\end{deluxetable}

\subsection{Ancillary Data}
\label{sec:supporting}
Beside {\it AstroSat} FUV and NUV observations, we also use several data sets from the literature and from the PHANGS-ALMA and PHANGS-MUSE surveys, which we summarize below.

\subsubsection{PHANGS-ALMA}
We used CO\,(2-1) emission line maps from the PHANGS–ALMA project \citep{Leroy2021}. The emission lines were observed at the rest frequency of $\nu=230.538$~GHz (Band 6), where the combination of receiver sensitivity and atmospheric transparency makes ALMA most efficient for $1''$ scale mapping of molecular gas. The ALMA data set provides a unique, $\sim$1\arcsec (100~pc at 20\,Mpc) view of nearby main-sequence galaxies at GMCs scale. 
The median of 1$\sigma$ noise in the cube of all targets is estimated to be 6.2 mJy beam$^{-1}$ at native resolution (corresponding to a brightness temperature of $\sim 0.17$~K). The velocity resolution of 2.5 km~s$^{-1}$ is chosen to improve the signal-to-noise ratio and quality of the deconvolution process. These observations were conducted using a bandwidth of 937.5\, MHz with a channel width of $\Delta \nu = 244$\,kHz. In contrast to UV observations from {\it AstroSat}, ALMA maps only cover the infrared-bright portion of the inner disk ($R_\mathrm{gal}=5-6\,$kpc) of our targets. Final data products are released using two different signal-selection masks, both indicating the likely location of real CO emission inside each cube. In this study, we used high completeness “broad” mask which is better for the detection of diffuse emission from the galaxy. The broad mask is built by convolving the cube into several lower resolution cubes and then a mask was created. Along with zero-moment maps that we used in this study, the uncertainty maps are also provided based on a three-dimensional noise model \citep[][, their section 7.2]{Leroy2021}.

\subsubsection{PHANGS-MUSE}
We also utilized products from PHANGS-MUSE (program IDs: 1100.B-0651/PI: E. Schinnerer; 095.C-0473/PI: G. Blanc; and 094.C-0623/PI: K. Kreckel). This sample encompasses 19 nearby star-forming spiral galaxies, as described by \cite{phangs-muse}. Out of these, 15 targets are also included in the PHANGS-{\it AstroSat} observations. The entire program took a total telescope time of $\sim$ 172 hours and each target has been observed with 3 to 15 pointings to cover the disk of the galaxy. Observations were conducted using MUSE wide-field mode (WFM; FoV = 1\arcmin), using the nominal (non-extended) wavelength from 4800 to 9300 \AA. The spectral resolution varies with wavelength but is typically around 2.75 \AA~(Full Width at Half Maximum, FWHM). The median Point Spread Function of the sample has a 0.69\arcsec\, FWHM at 6483.5 \AA. We use the emission line maps from \citet{phangs-muse}, which are created through the MUSE Data Analysis Pipeline (DAP) that fits each emission line along with the properties of the underlying stellar continuum. In this work, we use the H$\alpha$ and H$\beta$ emission lines as well as an estimate of the light-weighted stellar population ages determined from fits to the stellar spectra \citep{phangs-muse}.

\subsubsection{Literature data sets}    
Finally, we use the infrared and UV maps of our targets provided by \citet{z0mgs}, who supply an atlas of WISE (3.4$\mu$m to 22$\mu$m) and GALEX UV maps of $\sim$ 15000 nearby galaxies (D $\le$ 50\,Mpc). These GALEX maps are background-subtracted for UV background emission and corrected for foreground Galactic extinction. In our quality assessment analysis, we employ GALEX maps which are convolved to an angular resolution of 7\farcs5, in accordance with the \citep{z0mgs} dataset.

\begin{figure*}
    \centering
   \includegraphics[width=0.95\textwidth]{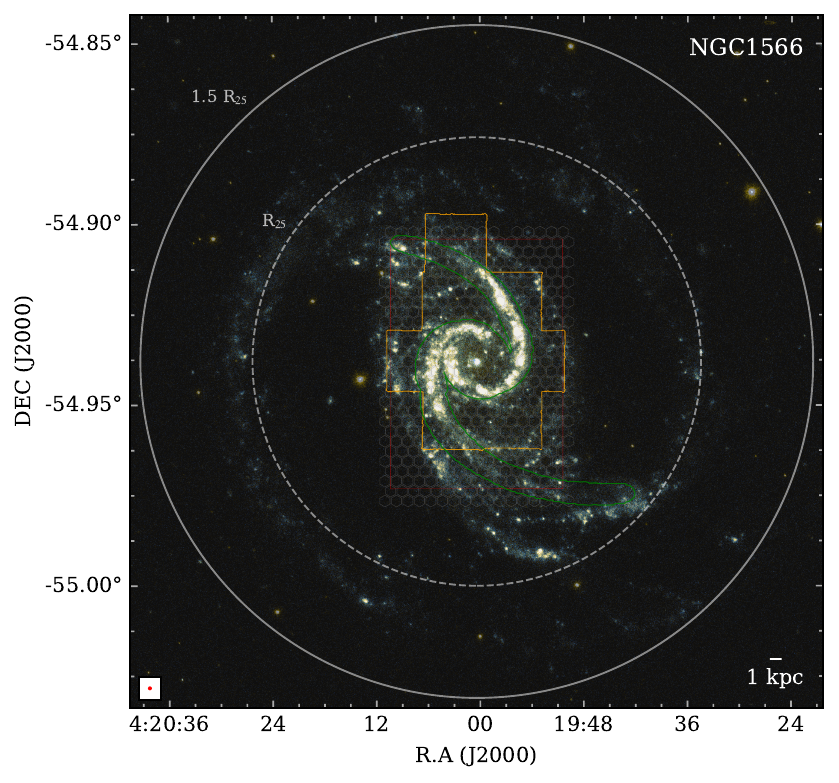}
\caption{{\it AstroSat} UVIT F148W (FUV) map (blue) and N263M  (NUV) (yellow) of NGC~1566 using an asinh brightness scale. The dynamical range of the N263M image is same as for the F148W map but normalized to the F148W/N263M central wavelength ratio for consistent cross-filter comparison. We show circles with radii of $R_{25}$ (dashed line) and 1.5 $R_{25}$ (solid line). The classification of spiral arms environments (green) is shown on the map. We show the PSF FWHM in the bottom left corner of the map as a red dot and a physical scale of 1kpc in bottom right corner. In addition, we plotted the 1-kpc size hexagonal apertures that we use to measure several quantities in sections \ref{sec:clumping} and \ref{sec:fuvha}. We overlaid the MUSE/VLT and ALMA footprints in orange and red colors respectively.}
    \label{fig:NGC1566}
\end{figure*}

\section{AstroSat Data Processing} \label{sec:procs}
\begin{figure*}
       \centering
       \includegraphics[width=0.75\textwidth]{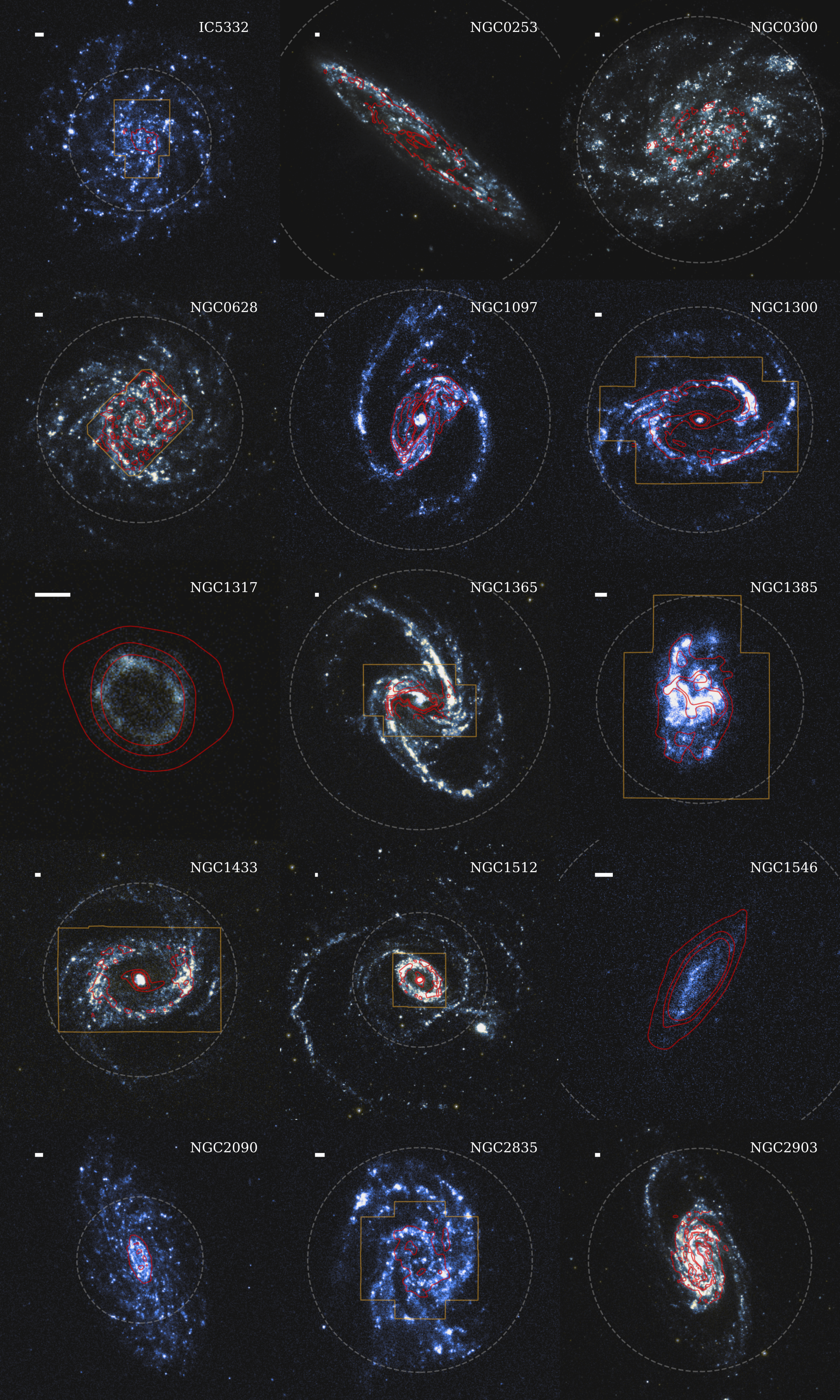}
\caption{Maps of PHANGS-{\it AstroSat} galaxies in the FUV band (blue) and any available NUV band (yellow) on an asinh brightness scale.  The dynamical range of NUV is same as FUV map but normalized to FUV/NUV central wavelength ratio for consistent cross-filter comparison. Red contours are from ALMA CO(2-1) data from 1 to 10 K~km~s$^{-1}$.  Both FUV and CO data are at a common 1.8$\arcsec$ resolution. Galaxies vary in angular and physical size. A radius of $R_{25}$ is shown as a dashed-line circle. The white line on the top left of maps have a length of 1\,kpc. We overlay the MUSE/VLT footprints with an orange color.}
    \label{fig:rgb_all_1}
\end{figure*}

\begin{figure*}
    \centering
   \includegraphics[width=0.75\textwidth]{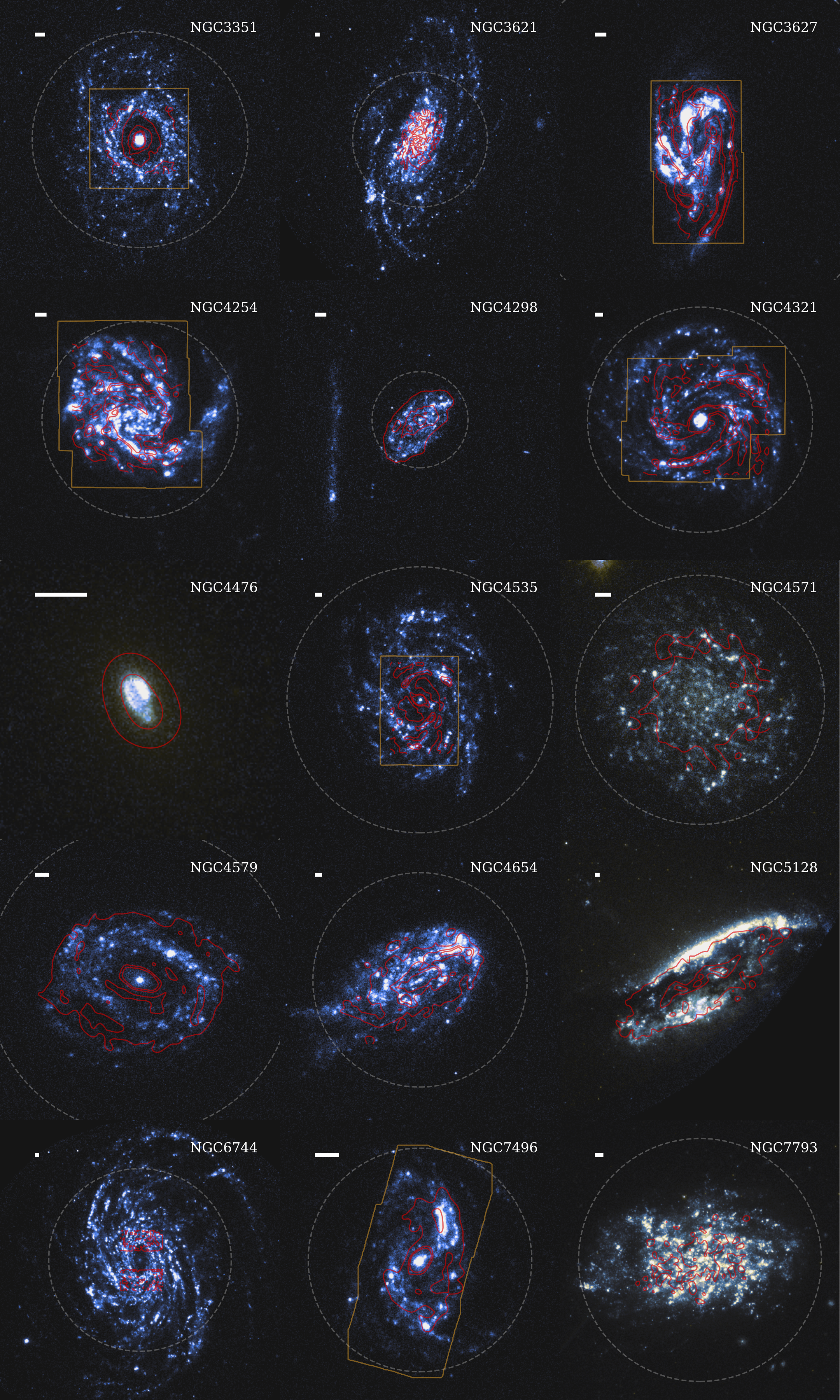}
\caption{Fig. \ref{fig:rgb_all_1} continued. In the case of NGC~0300, NGC~5128, NGC~0253, and NGC~7793 we use lower resolution ALMA CO data (TP and 7-m; $\sim 8\arcsec$) with contours between 0.75 to 10, 5 to 100, 5 to 100, and 1 to 10 K~km~s$^{-1}$ respectively.}
    \label{fig:rgb_all_2}
\end{figure*}

The Level-1 data products were generated using v2.1 and v2.2 of the {\it AstroSat} pipeline depending from \cite{Ravishankar} on the data set, and the Level-2 data were processed with the \textsc{CCDlab} software package \citep{postma17}. The pipeline accounts for various instrumental effects such as spacecraft drifts, jitter, and  thermal effects. A bright white dwarf (HZ4) was adopted as a photometric calibration source. A detailed set of criteria for selecting this source is discussed by \cite{tandon17}. The intensity of images are converted from observed counts/s (CPS) to flux density per 0.42\arcsec\ size pixel  ($F_{\lambda}$) using the updated in-orbit calibrations, following:
\begin{equation}
F_{\lambda} (\mathrm{ergs}\, \mathrm{s}^{-1}\,\mathrm{cm}^{-2}\,\mbox{\AA}^{-1}) = \mathrm{CPS} \times \mathrm{UC},
\end{equation}
where UC is the unit conversion factor derived from the zero-point (ZP) magnitude in the AB system as \citep[][their table 3]{tandon20}:
\begin{equation}
\mathrm{UC} = 10^{-0.4 \frac{(\mathrm{ZP}+2.407)}{\lambda_{\mathrm{mean}}^{2} } }.
\end{equation}
The $\lambda_{\mathrm{mean}}$ corresponds to the mean wavelengths of the UVIT filters in \AA. The uncertainties in $F_{\lambda}$ for different filters are due to the Poisson noise of the photon-counter detector as well as to uncertainties in the UC factors \citep{Subramaniam_2016}. In this study, we only propagate the uncertainty in CPS as the UC errors are taken to be small. These uncertainties yield $5\%$ uncertainty in the flux density for each filter \citep{Singh2021}. We align each image along the cardinal directions by reprojecting with the \textsc{reproject} python package\footnote{\url{https://reproject.readthedocs.io/en/stable/}}. Because the Point Spread Function (PSF) is minimally-sampled, we found that the adaptive resampling method was sufficient to maintain flux conservation in the reprojected image. Other interpolation and flux conserving methods produced a Moir\'e artifact pattern in the reprojected image, which was also noted by \citet{ClarkVerstocken2018A&A...609A..37C}, where they increased the pixel size in the reprojected image to avoid this artifact. Using the adaptive resampling method in \textsc{reproject}, we can maintain the same pixel scales as the original images. The final images have a pixel size of $0.42\arcsec$. As an example, we show a large panel of NGC~1566 FUV (F148W) map with environmental masks adopted from \cite{mask} in different colors and 1-kpc size hexagonal apertures in Fig \ref{fig:NGC1566}. We show the rest of PHANGS-{\it AstroSat} targets at FUV band in Figures \ref{fig:rgb_all_1} and \ref{fig:rgb_all_2} with contours from CO(2-1) molecular gas. In this study, we refer to FUV data specifically as data from the F148W band, which is the most available filter. Whenever the F148W band data is not available, we use other bands. Most galaxies have observations in the F148W filter, except NGC~1433, NGC~1512, NGC~4321, NGC~4571, NGC~4476, NGC~4579, and NGC~4254 which are in the F154W filter. In the case of NGC~0253, the only FUV band available is F169M.

\subsection{Background subtraction} 
\label{sec:background}

\begin{figure*}[t!]
    \centering
\includegraphics[width=0.96\textwidth]{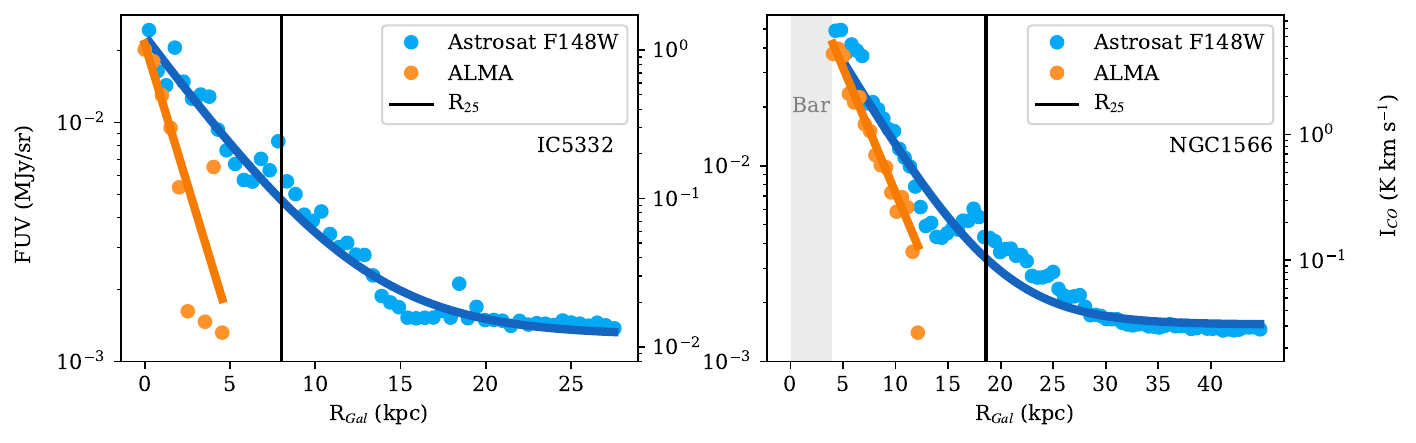}
\caption{Radial profiles of FUV (blue) and CO (orange) emissions for IC5332 (left) and NGC~1566 (right). The FUV radial profiles are presented before background subtraction and correction for Milky Way extinction. The black vertical line indicates the position of $R_{25}$ in kpc, while the gray line represents the maximum length of the bar, as reported by \cite{bar}, which we have excluded from the radial profiles. The solid lines show the exponential fits to the radial profiles as described in the text.}

    \label{fig:radial_profile1}
\end{figure*}

FUV imaging shows a near-constant flux of emission from ``blank'' sky, which we refer to as the {\it background} even though much of this emission may be from Galactic and solar system sources \citep{Kulkarni}. Taking advantage of the large field of view (FoV) of {\it AstroSat} observations, we have a radial profile of each object that extends beyond $\sim$30\,kpc galactocentric distance for most of the targets. We estimate the background by assuming the galaxy emission profile is an exponential disk \citep[e.g.,][]{Regan_2001} and include a constant background term:
\begin{equation}
\label{eqn:model}
S_{\lambda} = S_{0}\, e^{(-R_\mathrm{gal} /\gamma)} + B_{\lambda},
\end{equation}
where $S_{\lambda}$ is the emission at different bands, $S_{0}$ is the surface brightness at galactocentric radius $R_\mathrm{gal}=0$, $\gamma$ is the scale length, and $B_{\lambda}$ is the background value. We estimate background emission ($B_{\lambda}$) for all UV bands of {\it AstroSat}. The calibrated units of FUV and NUV maps are in erg~s$^{-1}$~cm$^{-2}$~\AA$^{-1}$~pix$^{-1}$, but we present maps scaled to units of MJy~sr$^{-1}$ to facilitate comparison to multiwavelength data. For comparison with the UV emission, we apply the same exponential model to the CO emission, fixing $B=0$, where we use the CO ``broad mask'' integrated intensity maps from \cite{Leroy2021}. This fit to the molecular line emission derives its scale-length to compare it with the FUV scale-length. We return to the comparison of the FUV-CO scale-lengths in more detail in Section \ref{sec:xuv} but here the primary use is to set the background level for the UV emission.

\begin{deluxetable}{lcccc}
\setlength{\tabcolsep}{0.8mm}
\tablecaption{Radial profile properties of galaxies.\label{tab:bkg-scale}}
\tablewidth{0pt}
\tablehead{
\colhead{Galaxy} &
\colhead{$E(B-V)$} &
\colhead{$B_\mathrm{FUV}$} &
\colhead{${\gamma}_\mathrm{FUV}$} &
\colhead{${\gamma}_\mathrm{CO}$	} \\
\colhead{} & \colhead{(mag)} & \colhead{($10^{-18}$~erg/s/cm$^{2}$/\AA)} & \colhead{(kpc)} & \colhead{(kpc)}
}
\startdata
IC5332 & 0.02 & 0.07 & $ 4.3 \pm 0.2 $ & $ 1.1 \pm 0.1 $ \\
NGC0300* & 0.01 & 0.08 & \nodata & $ 1.4 \pm 0.5 $ \\
NGC0628 & 0.07 & 0.07 & $ 5.2 \pm 0.2 $ & $ 1.9 \pm 0.2 $ \\
NGC1097 & 0.03 & 0.06 & $ 6.5 \pm 0.7 $ & $ 1.9 \pm 0.3 $ \\
NGC1300 & 0.03 & 0.08 & $ 4.6 \pm 0.3 $ & $ 2.0 \pm 0.6 $ \\
NGC1317 & 0.02 & 0.08 & $ 0.8 \pm 0.1 $ & $ 0.6 \pm 0.1 $ \\
NGC1365 & 0.02 & 0.05 & $ 8.3 \pm 0.4 $ & \nodata \\
NGC1385 & 0.02 & 0.09 & $ 1.5 \pm 0.1 $ & $ 1.2 \pm 0.1 $ \\
NGC1546 & 0.01 & 0.08 & $ 1.8 \pm 0.1 $ & \nodata \\
NGC1566 & 0.01 & 0.09 & $ 4.7 \pm 0.2 $ & $ 2.2 \pm 0.2 $ \\
NGC2090 & 0.04 & 0.15 & $ 5.6 \pm 0.4 $ & $ 1.9 \pm 0.6 $ \\
NGC2835 & 0.1 & 0.12 & $ 5.0 \pm 0.5 $ & $ 2.1 \pm 0.3 $ \\
NGC2903 & 0.03 & 0.07 & $ 4.0 \pm 0.2 $ & $ 1.6 \pm 0.1 $ \\
NGC3351 & 0.03 & 0.08 & $ 2.7 \pm 0.1 $ & $ 1.5 \pm 0.3 $ \\
NGC3621* & 0.08 & 0.12 & \nodata & $ 1.8 \pm 0.2 $ \\
NGC3627 & 0.04 & 0.07 & $ 1.8 \pm 0.1 $ & $ 1.9 \pm 0.2 $ \\
NGC4298 & 0.04 & 0.07 & $ 1.9 \pm 0.1 $ & $ 1.3 \pm 0.2 $ \\
NGC4535 & 0.02 & 0.06 & $ 4.0 \pm 0.2 $ & $ 2.7 \pm 0.5 $ \\
NGC4654 & 0.03 & 0.07 & $ 3.6 \pm 0.1 $ & $ 2.5 \pm 0.1 $ \\
NGC5128* & 0.1 & 0.18 & \nodata & $ 0.8 \pm 0.0 $ \\
NGC6744* & 0.04 & 0.13 &  \nodata  & $ 3.6 \pm 0.4 $ \\
NGC7793 & 0.02 & 0.05 & $ 1.4 \pm 0.1 $ & $ 1.2 \pm 0.1 $ \\
NGC7496 & 0.01 & 0.07 & $ 1.7 \pm 0.1 $ & $ 0.9 \pm 0.1 $ 
\enddata
\tablecomments{Second column: foreground color excess from \cite{sfd98}. Third column: background emission at F148W. Fourth column: scale length from F148W band. Fifth column: scale length from CO. \\
* The background ($B_\mathrm{FUV}$) is obtained from ``mean-level'' method for these targets, described in Section~\ref{sec:background} and no scale length is determined.}
\end{deluxetable}

To minimize the impact of the FUV bright central bump commonly found in galaxies, we exclude the bar region from the radial profile by using a radius of $1.3\times R_{\mathrm{Bar}}$, where $R_\mathrm{Bar}$ is adopted from \citet{bar} \citep[see also ][]{mask}. For NGC~4535 and NGC~6744, we adopt bar sizes of 5.1 and 4\,kpc, respectively. We determine the radial profiles by averaging the flux in annuli with widths of 500\,pc from the center of the galaxy up to a maximum galactocentric distance of $<50$\,kpc. This maximum distance is chosen through visual inspection of different galaxies to exclude any bright nearby sources. Figure \ref{fig:radial_profile1} presents the radial profiles of {\it AstroSat} observations in the FUV band for IC~5332 and NGC~1566. The remaining radial profiles are presented in Appendix \ref{ap1}. In the case of NGC~0253, we use the only available filter in the FUV band: F169M.

For most of our targets, {\it AstroSat}'s large ($28'$) field of view (FoV) includes enough sky area at larger galactocentric distances that our simple exponential model (Eq.\ref{eqn:model}) is sufficient for determining the background level of images robustly. However, some targets have relatively smaller amounts of empty emission-free sky due to the small coverage of observations or several bright sources in the FoV. In these cases, regression over the exponential model fails for a few targets. For example, NGC~0300 is sufficiently nearby that we only have a radial profile out to a small galactocentric distance of $R_\mathrm{gal}<12$\,kpc. Furthermore, the exponential model is not appropriate for the highly-inclined NGC~0253, where we see many bright gas clumps at $R_\mathrm{gal}<10$\,kpc. In the cases of NGC~3621 and NGC~5128, we decided not to use the regression to determine the background level due to a number of nearby bright sources. Finally, in the interacting system of NGC~1512-1510, a simple exponential model could not describe the complicated radial profile of this merging system, hence the background value could not be estimated robustly from radial profiles.

In these cases, we instead used a $\textit{``mean-level''}$ estimate of the backgrounds.  The background value for each of these targets is estimated by taking the mean value of an annulus around each object. The annulus radius is provided by visual inspection, ensuring the exclusion of nearby sources, mostly $\mathrm{R}>3\,\mathrm{R}_{25}$. However, these annuli values might still suffer from bright stars or reaching the edge of the FoV. We reject bright sources using Chauvenet's criterion in the annulus data distribution (i.e., iteratively rejecting data $>3\sigma$ where $\sigma$ is the standard deviation and then recalculating $\sigma$). The standard deviation is inserted into the Chauvenet criterion using the median absolute deviation (MAD) as an estimator. The MAD standard deviation is provided by the \texttt{astropy.stats} module\footnote{https://docs.astropy.org/en/stable/stats/index.html}. Although the MAD standard deviation is robust to outliers, it could be zero or much smaller than the normal standard deviation due to the faint nature of background emission. In that case, we used ordinary standard deviation to avoid failure of the Chauvenet criterion.  After rejecting data using this criterion, we estimate the background as the mean in the annulus. Regarding the discussion of the new XUV disk galaxy NGC 6744, which has spiral arms extending beyond 2 $R_{25}$ (as detailed in Section \ref{sec:xuv}), we have decided to avoid using the radial profile or $\textit{``mean-level''}$ background. The radial profile of this galaxy is not exponential and the estimation of the background is not reliable due to the presence of numerous bright stellar clusters and \ion{H}{2} regions. Therefore, we have carefully chosen a few 1.5\arcmin\, apertures that are free from bright sources, and estimated the mean emission from those locations. The estimated background is 20 and 34 percent higher than $\textit{``mean-level''}$ and radial profile respectively.

Table \ref{tab:bkg-scale} lists the foreground extinction, background values, and scale lengths of our sample for both FUV band and CO emission.

\begin{figure*}
    \centering
   \includegraphics[width=0.95\textwidth]{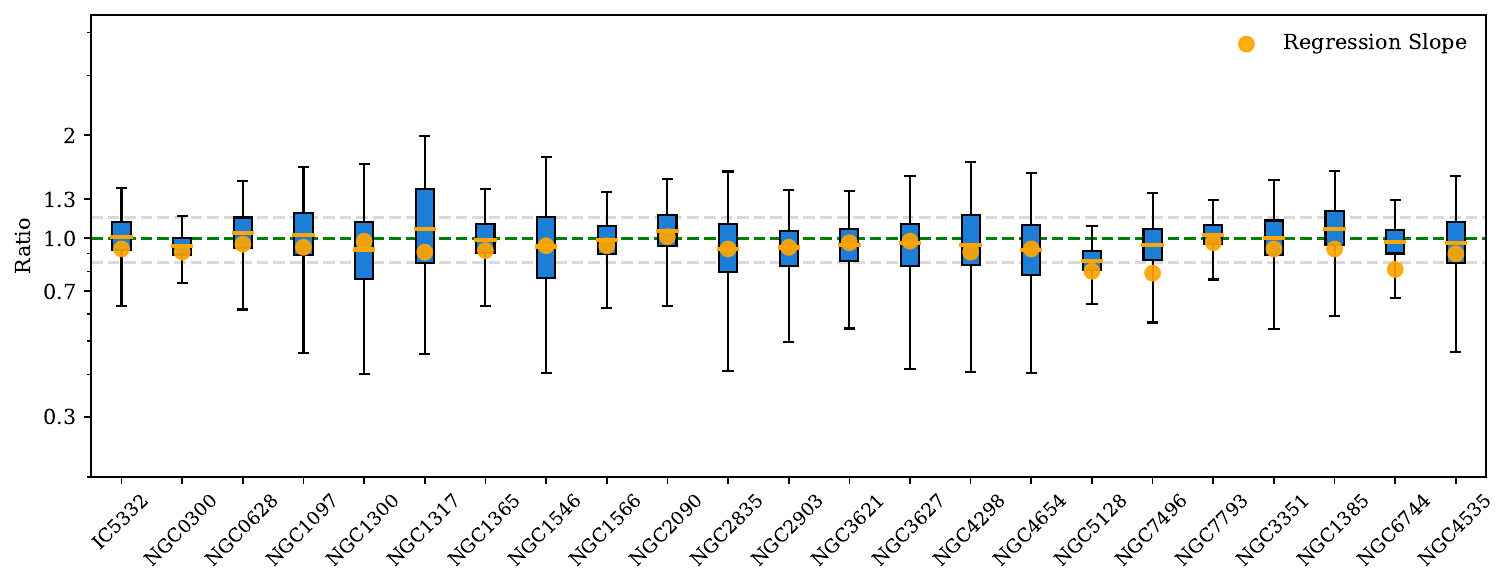}
\caption{Pixel value distribution of {\it AstroSat} F148W to GALEX FUV maps at a common beam of 7\farcs5 for different galaxies. The $y$-axis is plotted in a log scale. Both maps are background subtracted and corrected for Milky Way foreground extinction using the same approach. The green dashed line is the ratio of unity and the gray horizontal lines show a confidence band of 15 per cent. The slope of regression between pixel values of GALEX and {\it AstroSat} maps (described in Sec. \ref{sec:galex}) are shown as orange points on box plots.}
    \label{fig:fuv_galex}
\end{figure*}

\begin{figure}
    \centering
    \includegraphics[width=0.95\columnwidth]{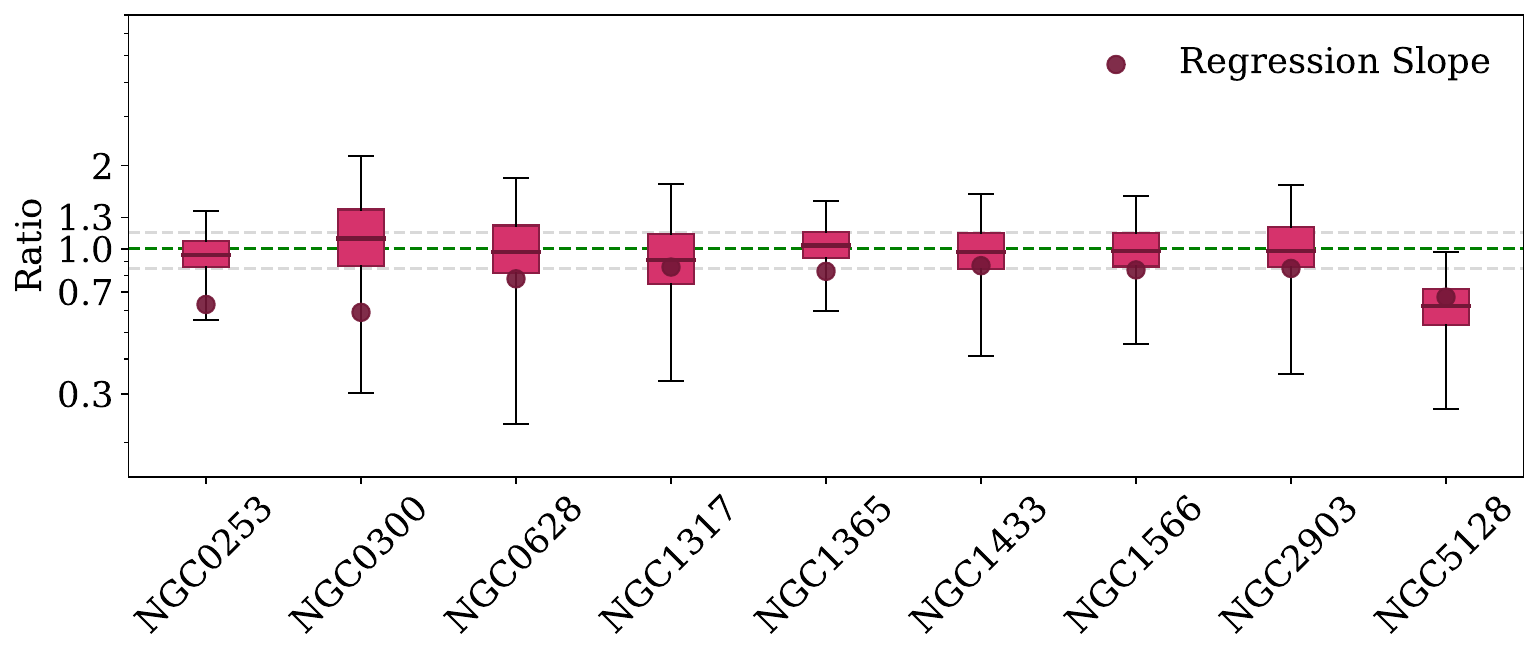}
\caption{Same as Fig. \ref{fig:fuv_galex}, but for {\it AstroSat} N219M to GALEX NUV maps.}
    \label{fig:nuv_galex}
\end{figure}

\subsection{Foreground extinction} 
As Milky Way foreground dust extincts UV emission of each target, we first corrected both {\it AstroSat} FUV and NUV fluxes for foreground extinction following the \citet{z0mgs} procedure. A uniform foreground dust distribution is expected due to the small angular size of each target. The ratio of total to selective extinction $R_{\mathrm{FUV}}$ and $R_{\mathrm{NUV}}$ is given by \citet{peek2013} as a function of color excess $E(B-V)$. The extinction is estimated using $A_{\lambda} = R_{\lambda}\, E(B-V)$ at the FUV and NUV wavelengths. Based on \citet{sfd98} dust maps and its $E(B-V)$ value for each galaxy, $A_{\mathrm{FUV}}$ and A$_{\mathrm{NUV}}$ is estimated for each target in the sample, resulting in a median foreground of A$_{\mathrm{FUV}}=0.21$ mag but the correction shows $A_{\mathrm{FUV}}>1$ mag for NGC\,5128 and NGC\,2835.

\subsection{Quality Assurance}
We characterize the quality of the {\it AstroSat} imaging through three separate checks: we assess the resolution by fitting point sources in the data, we validate the alignment against higher resolution HST data, and we compare the flux scales to archival GALEX maps from \cite{z0mgs}.  In all cases, the {\it AstroSat} data meet acceptable quality standards.

\subsubsection{Resolution and Alignment}

We characterize the point spread function (PSF) of the data reduction empirically using point sources in the final reduced products.  The PSF of the final data is primarily set by the quality of the tracking solution and alignment of subimages, which can vary based on the brightness of sources in the field and number of contributing observations to a final image.

To measure the PSF, we identify point sources in the image, stack the images, and fit a Moffat PSF model \citep{MoffatProfile} to the resulting image stack.  We adopt the Moffat PSF model since it appears to characterize the PSF of bright sources well, including a compact core with extended wings that result from deviations in star tracking. We develop a customized solution to PSF characterization because of the relatively low number of photon counts in an individual source and the relatively small number of stars that are bright in the UV.  First, we smooth the image with a Gaussian kernel with a standard deviation of 2 pixels. We then identify point sources using the DAO Star Finder algorithm \citep{Stetson87} as implemented in the \textsc{photutils} package \citep{photutils} in \textsc{python}. Of these candidate sources, we cross match to objects identified in Gaia Data Release 3 \citep{gaia2}, retaining only those sources within $2''$ of a Gaia DR3 source with a parallax/error value of 10 or more.  Such sources are  likely to be stars in our own Galaxy.  These selections yielded typically 20-40 sources per field.  We stack the extracted sources on the maximum brightness pixel to form a single representative star image and fit a two-dimensional Moffat profile to the resulting stack.  The median resolution of the FUV bands is $1.4''$ and the median resolution of the NUV bands is $1.2''$. The better resolution for the NUV bands comes from the improved tracking of the higher signal-to-noise values in the NUV images \citep{tandon17}. The derived PSF parameters are recorded in header of each image in the data release.

We use the astrometric solution implemented in \textsc{CCDLAB}, following the approach outlined in \citet{postma20}.  This method identifies UV-bright Gaia sources in the field and fits an astrometric solution.  We evaluate the quality of the solution by comparing NUV data (where available) to the F275W filter from the {\it Hubble} Space Telescope imaging \citep{phangs-hst} through cross correlation.  In all cases, we find better than $0.4''$ accuracy in the astrometric alignment.

\begin{figure}
    \centering
   \includegraphics[width=0.97\columnwidth]{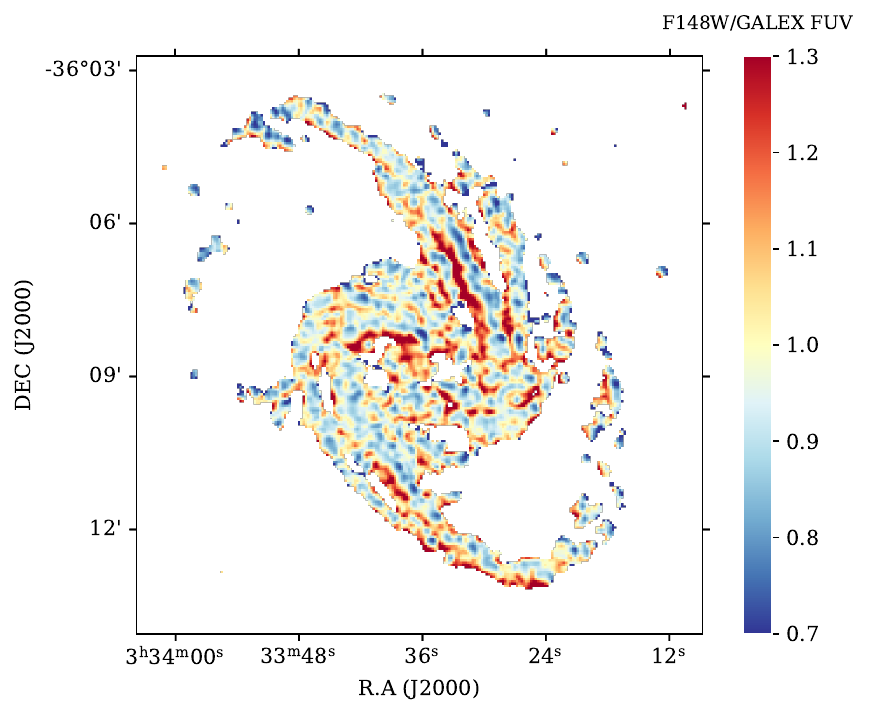}
        \caption{Ratio of {\it AstroSat} (F148W) to GALEX (FUV) in NGC~1365 at a common beam of 7\farcs5. The color bar at right shows the ratio values in a linear scale. We expect a ratio $<1$ (yellow to blue colors) as the central wavelength of the F148W filter is shorter than the GALEX band. As we discuss in Sec.\ \ref{sec:galex}, the slope of regression between  {\it AstroSat} and GALEX $\beta=0.92$ with a offset of 0.001\,MJy/sr.}
    \label{fig:ratio_map}
\end{figure}
\subsubsection{Comparison with GALEX}\label{sec:galex}

To validate our reduction, we performed a comparison with GALEX observations. Only a portion of {\it AstroSat} data is comparable with GALEX bands, notably F148W and F154W in {\it AstroSat} can be compared to the GALEX FUV band, and N219M and N242W from {\it AstroSat} can be compared to the NUV band. Even these comparisons should be made with caution due to differences between the widths of filters and their central wavelengths.

The units of {\it AstroSat} maps are converted to MJy/sr and then we convolve the {\it AstroSat} images to a common resolution of 7.5\arcsec\, and reproject the {\it AstroSat} images to match the GALEX observations from \cite{z0mgs}. Pixel values below 5$\sigma$ RMS are masked using the \textsc{astropy.stats.sigma-clipped-stats} module on GALEX maps. Figures~\ref{fig:fuv_galex} and \ref{fig:nuv_galex} show the pixel value distributions of the {\it AstroSat} to GALEX ratio maps, indicating 25th- and 75th-percentiles of distribution as the top and bottom of the boxes as well as the median value (center line). The lines extending from the box indicate the minimum and maximum of the the distribution. Using \textsc{sklearn.linear\_model} module\footnote{https://scikit-learn.org/stable/modules/generated/sklearn.linear-model.LinearRegression.html} Ordinary Least Squares (OLS) regression, we also fit a line to average of every 30 pixel values, using $Y (\mathrm{Astrosat})=\beta\,X(\mathrm{GALEX})+C$ over data in linear space to obtain a slope to charaterize the linear relation between data values and quantify any possible constant offsets between the data. After masking values below three $\sigma$ for both {\it AstroSat} and GALEX maps at a common resolution, we found good agreement between GALEX FUV and {\it AstroSat} F148W bands with a median of $\beta=0.92$ and with an offset of $C=0.001$\,MJy/sr which is lower than the noise value by a factor of two in most targets. We expect a systematically lower fraction due to differences in the passband and central wavelength of F148W compared to the GALEX FUV filter ($\lambda_\mathrm{central}$=1528\AA). In the NUV, we found a median of $\beta=0.83$ and $C=0.002$\,MJy/sr among targets at N219M band, which is reasonable given the difference with the GALEX NUV filter bandwidth and the central wavelength. We plot $\beta$ values over ratio of pixel value distribution for each galaxy in Fig. \ref{fig:fuv_galex}, \ref{fig:nuv_galex} at FUV and NUV bands, respectively. Fig. \ref{fig:ratio_map} shows the ratio map of {\it AstroSat} F148W to GALEX FUV in NGC~1365 on a linear scale where we see a good agreement at the 10 percent level across the galaxy.

\begin{figure*}[!t]
    \centering
    \includegraphics[width=0.8\textwidth]{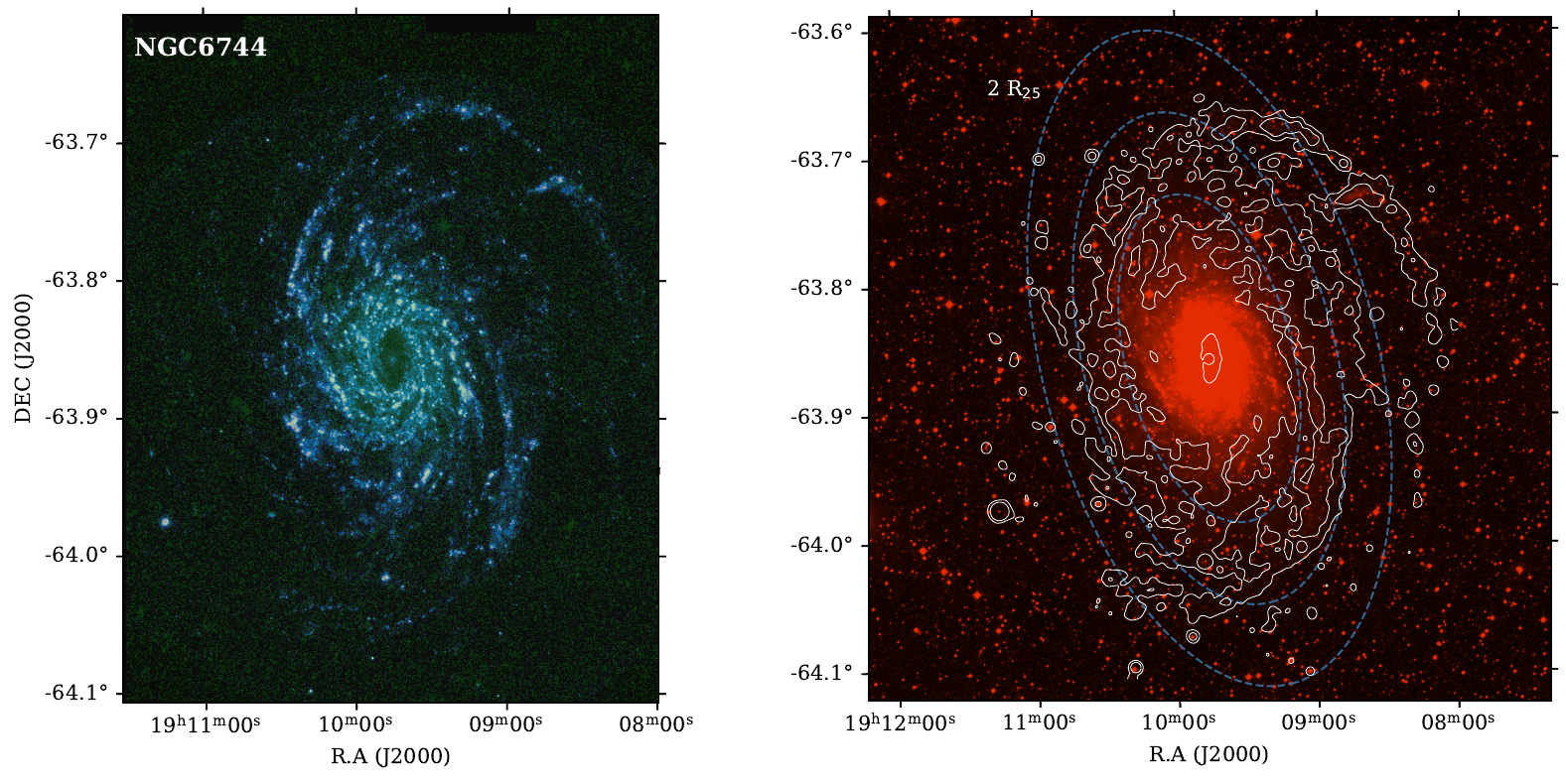}
    \includegraphics[width=0.8\textwidth]{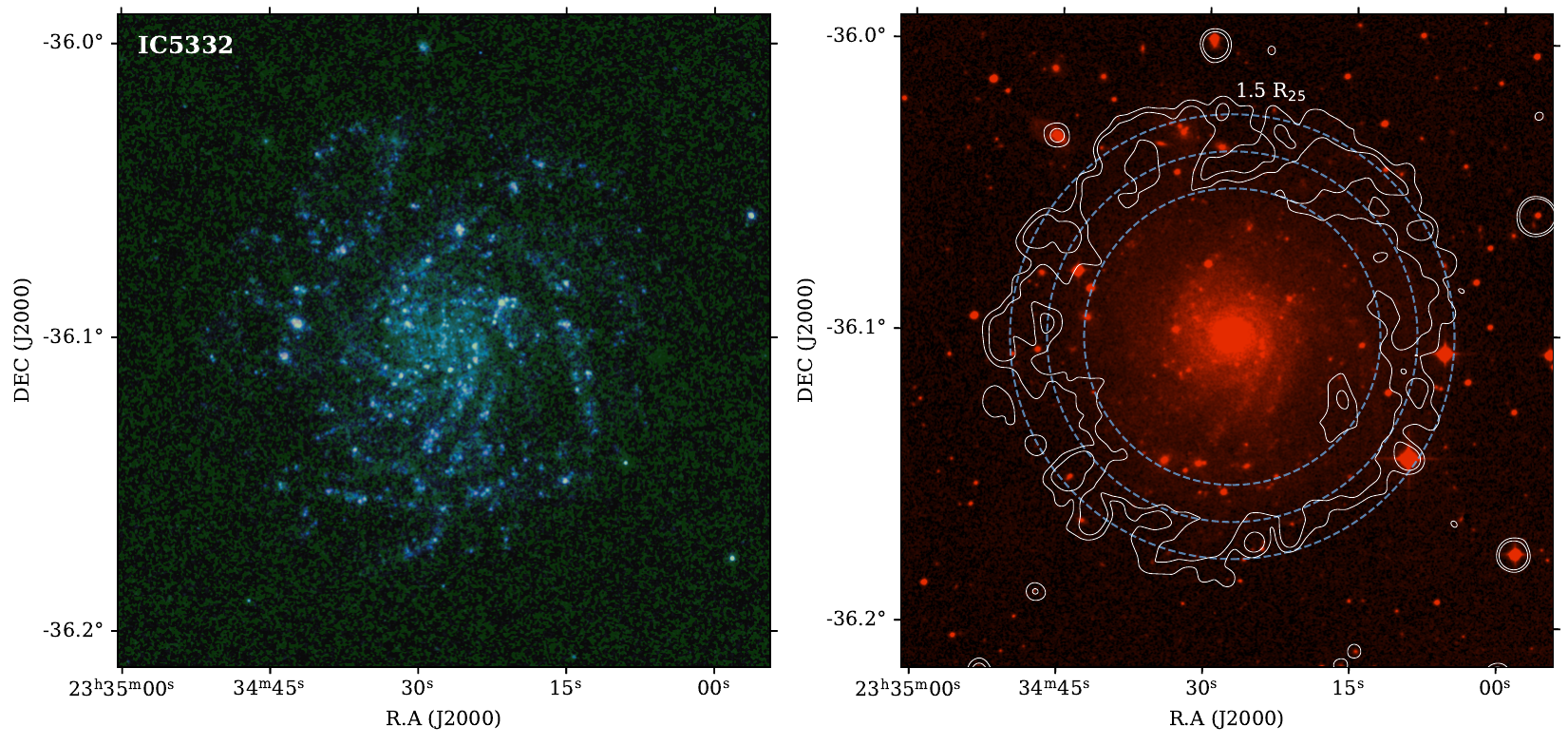}
    \caption{New candidates for XUV-disks galaxies. From top to bottom panels: NGC~6744 and IC~5332 Left: Maps of the {\it AstroSat} (F148W) in blue and 2MASS (K$_{s}$) in green at 2MASS resolution. The FUV map is in a asinh brightness scale. Right: Digitized Sky Survey (DSS) 2 red map overlaid with white contours of {\it AstroSat} F148W at levels of 26.2 and 27.9 $M_{\mathrm{AB}}$ for NGC~6744, 27.2 and 28 $M_{\mathrm{AB}}$ for IC~5332. The FUV maps are not corrected for internal extinction. Blue ellipses are $R_{25}$ with 0.5 interval for NGC~6744 and 0.25 for IC~5332.}
    \label{fig:XUV_disk_image}
\end{figure*}

\section{XUV disks}
\label{sec:xuv}

In the following sections, we illustrate the scientific potential of arcsecond resolution and relatively wide field-of-view of the FUV imaging in the context of the PHANGS multiwavelength data set. Unlike the PHANGS observations conducted by ALMA, MUSE, JWST, and HST, which are predominantly confined to the inner disks of galaxies, {\it AstroSat}'s observations extend significantly beyond the optical disks, offering a more expansive view of the low-density structure of galaxies \citep{phangsalma,phangs-muse,phangs-hst,lee2023}. Here, we use the {\it AstroSat} data to identify extended ultraviolet (XUV) disks beyond the optical disk of galaxies. \citet{Thilker2007} searched for the XUV disk galaxies in the local universe (D$<40$\,Mpc) using GALEX observations, finding UV-bright complexes beyond the optical radius in 30\% of spiral galaxies \citep{Thilker2007}. The detection of XUV emission is particularly intriguing as it implies the presence of recent massive star formation. Remarkably, this formation is taking place in poorly understood environments, characterized by low ISM pressure and low metallicity. These unconventional conditions foster extreme scenarios, which offer valuable opportunities to challenge and refine our current theories of star formation. Subsequent observations in the Local Volume and at a distance beyond 100~Mpc have further confirmed the ubiquity of XUV disk galaxies \citep{Lemonias2011}, reinforcing their importance in understanding galaxy evolution.

\subsection{Background}
There is some debate about what mechanisms power the XUV emission in galaxies, which could arise from different types of stars. Spectroscopic observations in the M83 and NGC4625 indicate that the emission of XUV disks ($>1.5 R_{25}$) is linked to the population of young stars associated with low-mass stars \citep{GildePaz2007}. \citet{GildePaz2007} highlighted that the emission lines in the XUV disks of these galaxies are formed by single stars with masses between 20-40 M$_{\odot}$. Furthermore, their findings indicate that the role of planetary nebulae and blue horizontal branch (HB) stars in shaping XUV disks should be negligible. The highly excited gas of planetary nebulae and its high electron temperatures cannot explain the observed spectroscopic emission lines in XUV disks. On the other hand, the blue color of NUV-optical emission, as well as the structured shape of XUV disks (e.g., spiral arms), are pieces of evidence that do not support the role of HB stars as candidates for powering XUV disks \citep{GildePaz2007}. Previous findings also confirm that FUV complexes with masses of $10^{3}-10^{4} \:{\rm M_\odot}$ are being ionized by single stars \citep{Gil2005}. On the other hand, UV emission could also originate from white dwarfs in XUV disks \citep{Sahu}.

Different scenarios are proposed for triggering the star formation process in XUV disks, such as interactions, perturbations, and gas accretion \citep{bush2008,Bush2010,Das}. In addition, spiral density waves could propagate from inner to outer disk and result in gravitationally unstable regions which eventually form stars \citep{Lemonias2011}. As one possible indication of the formation mechanisms, radio observations also find extended \ion{H}{1} disks associated with XUV emission in M~83 and NGC~4625 \citep{m83_HI,bigiel2010,Bush}.  In the case of the late type galaxy NGC~0628, a faint trail of UV emission extends into the HI outer disk, which resembles a spiral-like structure \citep{Das2020}. This significant amount of HI is maybe due to a low-efficiency process of turning neutral gas into stars \citep{GildePaz2007}. The consumption timescale of \ion{H}{1} in XUV disks is estimated to range from a few Gyr up to the Hubble time \citep{Gil2005}. Galaxies hosting XUV disks might indicate the presence of molecular clouds as well as extended \ion{H}{1} well beyond their stellar disks. The detection of faint CO emission in the outer disk is essential to understand the star formation efficiency (SFE) in this environment. For instance, \citet{Dessauges} detected CO(1-0) in M~63 with a low SFE, compared to inner disk regions using the IRAM 30~m telescope. \citet{koda22} used deep ALMA observations to identify small molecular clouds associated with the XUV emission. The mechanisms behind the formation of XUV disks in galaxies are still unclear and the number of well-studied targets remains small.  In this study, our objective is to use new UVIT observations to identify new potential candidates using the XUV disk classification scheme proposed by \citet{Thilker2007}.


\begin{figure}
    \centering
    \includegraphics[width=\columnwidth]{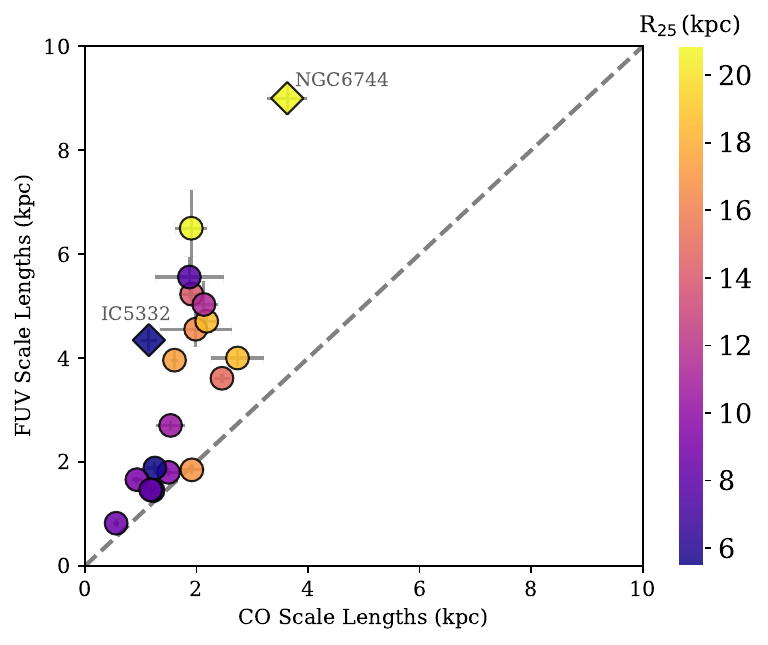}
    \caption{FUV F148W scale lengths versus CO scale lengths obtained from radial profiles of galaxies, color coded with $R_{25}$. We show two new XUV disks candidates with diamond markers.}
    \label{fig:scales}
\end{figure}

\subsection{New candidates}

Following the prescriptions of \citet{Thilker2007} and \citet{Lemonias2011}, we used a foreground-extinction-corrected, background-subtracted UV surface brightness ($\mu_\mathrm{FUV}$=27.25 AB magnitudes arcsec$^{-2}$) as a threshold level to identify FUV clumps in XUV disks. This UV surface brightness corresponds to $\Sigma_\mathrm{SFR}$= $3\times10^{-4}$~M$_{\odot}$~yr$^{-1}$~kpc$^{-2}$, which is based on a Salpeter IMF and calibrations from \citet{Kennicutt98}. This type of XUV disk requires more than one structured UV complex (e.g. spiral segments) and is known as a Type 1 XUV-disk or ``M~83 like'' object. We do not consider diffuse UV emission as a criterion for XUV disks, as it could be related to other recent SF processes or originated from hot core-helium-burning stars \citep{Hoopes2005}. On the other hand, Type 2 XUV disks are often dominated by a large, blue, low-surface-brightness outer area with a higher specific star formation rate and less structured emission \citep[see Section 3 in][]{Thilker2007}.

Using the radial profiles presented in Section \ref{sec:background}, we measured the scale lengths of FUV emission and CO molecular gas. Figure \ref{fig:scales} compares the scale lengths of these two different data sets. On average, the scale length of FUV ($\gamma_\mathrm{FUV} \simeq 3500$ pc) is about twice that of CO molecular gas. These results indicate that the molecular gas is concentrated in the stellar disk, while structured clumps of young massive stars, with a longer lifetime, can exist far beyond the optical disks. Significantly, as highlighted by \cite{Bigiel_2010_b}, it has been observed that the scale length of \ion{H}{1} is typically twice that of the FUV emission. Consequently, this leads to an extended depletion time for neutral gas, suggesting that XUV disk galaxies are frequently enveloped by extensive \ion{H}{1} gas reservoirs. It is important to emphasize that the PHANGS-ALMA observations are primarily focused on the optical disk regions of galaxies, where the majority of massive CO clouds (with masses exceeding 10$^4$ M$_{\odot}$) are typically found. This emphasis is underscored by prior detections of CO in the XUV disks, where such clumps are often faint and isolated. Consequently, incorporating observations of diffuse clouds, which would requires deeper observations, is unlikely to significantly impact the determination of CO scale lengths.

When comparing scale lengths, our focus was on identifying galaxies with large FUV scale lengths, particularly in comparison to CO emission. Among the galaxies in our sample, NGC~6744 possesses the largest scale length, with $\gamma_{\mathrm{FUV}}=9$\,kpc and $\gamma_{\mathrm{CO}}=3.8$\,kpc. We note that the radial profile of this galaxy is not exponential, which has resulted in poor regression results, and hence the uncertainty is likely underestimated. We also observe that the obtained parameters of this galaxy, such as FUV scale length, amplitude, and background, are close to being outside of the reasonable range that we confined our initial regression parameters to. This is another indication that our regression does not adequately address the radial profile properties of this galaxy. Therefore, we did not use the background measured from the radial profile for this galaxy, as also noted in Table \ref{tab:bkg-scale}. This high value FUV scale length is not surprising as $R_{25}$ is also more than 20\,kpc in this galaxy. 
We investigate all targets with large $\gamma_\mathrm{FUV}$ relative to $\gamma_\mathrm{CO}$ and identify NGC~6744 and IC~5332 as also having FUV bright segments structures beyond their optical disk, considering these as new candidates for XUV disks. We note that these targets are not covered in the previous studies of XUV-disk galaxies by \cite{Thilker2007}.  We show new XUV disk candidate galaxies in Fig.~\ref{fig:XUV_disk_image} and discuss the features of these XUV-disk galaxies below.

\textit{NGC~6744:} This galaxy is an intermediate gas-rich spiral galaxy that hosts several FUV-bright clumps arranged in a spiral-arm structure at a distance of approximately 29 kpc from its center. This galaxy is situated near the Galactic plane, which has led to it being less studied in surveys at other wavelengths due to the observational complexities of observing at low Galactic latitude. However, we have identified two clearly visible FUV-bright spiral structures, both located beyond 1.5 times the optical radius ($1.5 R_{25}$), without any corresponding bright optical counterparts. Interestingly, we do not see any bright $3.3$$\mu$m Polycyclic Aromatic Hydrocarbons (PAH) emission from nearby stellar population in this area, as traced by IRAC 3.6~$\mu$m images \citep[][and references therein]{6744_dust}. Low resolution \ion{H}{1} observations also confirmed the presence of these two spiral arms, suggesting streaming motions along the \ion{H}{1} arms as a result of interaction with companion galaxy \citep{6744}. The companion IB(s)m NGC~6744A galaxy is visible in the northwest region in the FUV band as well. Furthermore, GMOS/IFU observations in the center of this galaxy suggested two periods of star formation which the last one happened one billion years ago, due to a merger event \citep{6744_agn}. Having this evidence, we propose that NGC~6744 is a Type 1 XUV-disk and its nature is probably related to the accretion of gas from its companion in a merger event.

\textit{IC~5332:} this galaxy is also another intermediate spiral galaxy but with 5$\times$ lower stellar mass (M$=0.41\,\times\,10^{10}$ M$_{\odot}$) in comparison with NGC~6744. Although we see several star forming complexes beyond the optical disk, we do not identify any ordered spiral-like features. Unlike NGC~6744, there are several irregular clumpy structures at $R>R_{25}1.5$. \citet{Thilker2007} found that Type 2 XUV disks are more common in late-type/low-mass spirals, similar to IC~5332. Using 2MASS K$_{s}$ data from \cite{twomass}, we are unable to define a low surface brightness (LSB) zone for these irregular structures in the outer disk, as the K$_{s}$-band luminosity is faint in this region. Parkes and ATCA observations of IC~5332 also confirm the existence of an extended \ion{H}{1} emission in the outer disk \citep{5332}. On the other hand, using the H$\alpha$ emission map, we see a good match with FUV emission (but not with old stellar population sites traced by IRAC 3.6~$\mu$m) in several clumps located in the southeast and west regions. These results might suggest an ongoing recent star formation process that results in diffuse UV emission and not a outer spiral arm feature. This scenario might cast doubt on IC~5332 being an XUV-disk galaxy. Furthermore, being in the LGG~478 group of galaxies, makes IC~5332 a possible part of a merger event. A detailed study of star formation history in IC~5332 could shed light on understanding different episodes of star formation and confirm such a scenario. The origin of the IC 5332 XUV emission remains unclear, but this target remains a candidate for an XUV disk.

\section{Structural Measurements of FUV Emission}
\label{sec:clumping}

The motivation behind studying the clumping factor in CO and FUV emission is to understand the structure and distribution of molecular gas and newly born stars in galaxies. \cite{Leroy_2013} indicate that models describing the structure of the ISM require detailed information on the mapping between the observed average surface density, referred to as the ``area-weighted'' surface density, at a resolution of a kpc. This mapping plays a crucial role in comparing observations with theoretical models and enhancing our understanding of galaxy structure.

In this section, we explore the brightness distribution and emission structures of the FUV map as measured using the {\it clumping factor}.  We compare the measures for FUV emission with the results for the CO.  We contrast the clumping factor measurements with the results derived from the {\it Gini coefficient}, which can be used to characterize the structure of emission in a map \citep[e.g.,][]{Davis}.  Both the clumping factor and the Gini coefficient are scalar measurements of the shape of the probability density function.

\subsection{Clumping Factor}
\label{sec:clumpingg}
The clumping factor is a scalar measurement that traces the width of the probability density function relative to its mean, which can be interpreted to measure the smoothness of the emission. We define the clumping factor following \cite{Leroy_2013,Sun_2022}:
\begin{equation}
c\sbsc{pix,\,\theta\,pc} =
\frac{\left(\sum\limits_i I_{\textit{i},\,\theta\,pc}^2\right)N\sbsc{pix}}{\left(\sum\limits_i I_{\textit{i},\,\theta\,pc}\right)^2},
\label{eq:clumping}
\end{equation}
where $I_{\textit{i},\,\theta\,pc}$ is the surface brightness of region ($i$) at resolution of $\theta$ pc. The summation is over all pixels inside the region and $N_\mathrm{pix}$ is the total number of pixels inside that region. Eq.~\ref{eq:clumping} is the same definition from \cite{Leroy_2013}, where it is a ratio between the mass-weighted mean and the area-weighted mean of surface density.  The measurements from \citet{Leroy_2013} indicated that molecular gas is highly clumped ($c\gtrsim 7$), whereas atomic gas has a smooth distribution ($c\approx 1$). This result could be interpreted as showing that most of the 21\,cm emission originates from a diffuse medium with a high volume filling factor \citep{Leroy_2013} while most of the molecular emission is organized into individual clouds. However, the UV clumping factor has not been examined before. The photospheric emission from young stars, traced by the FUV, should have slightly different clumping structures compared to molecular gas since they are much brighter in the outer disk. Additionally, the FUV emission may originate from a wide range of stellar populations with different ages, so the clumping factor could lie between that of molecular and atomic gas. Here, we explore the clumpiness in CO versus FUV emission to show how the formation of massive stars may derive from the ISM clumpiness. Having 31 galaxies spanning a wide range of mass and metallicity, we can also inspect the clumpiness in different galactic environments. We restrict our analysis to the regions where FUV maps overlap with the ALMA CO observations and carry out the analysis at a common resolution of 175 pc.

\begin{figure*}
    \centering
    \includegraphics[width=0.99\columnwidth]{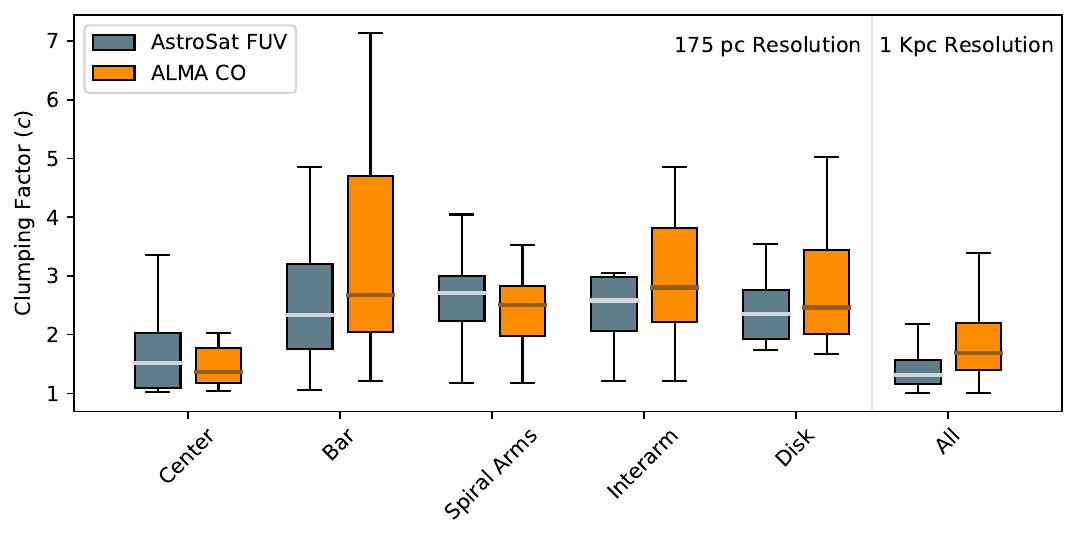}
    \includegraphics[width=0.99\columnwidth]{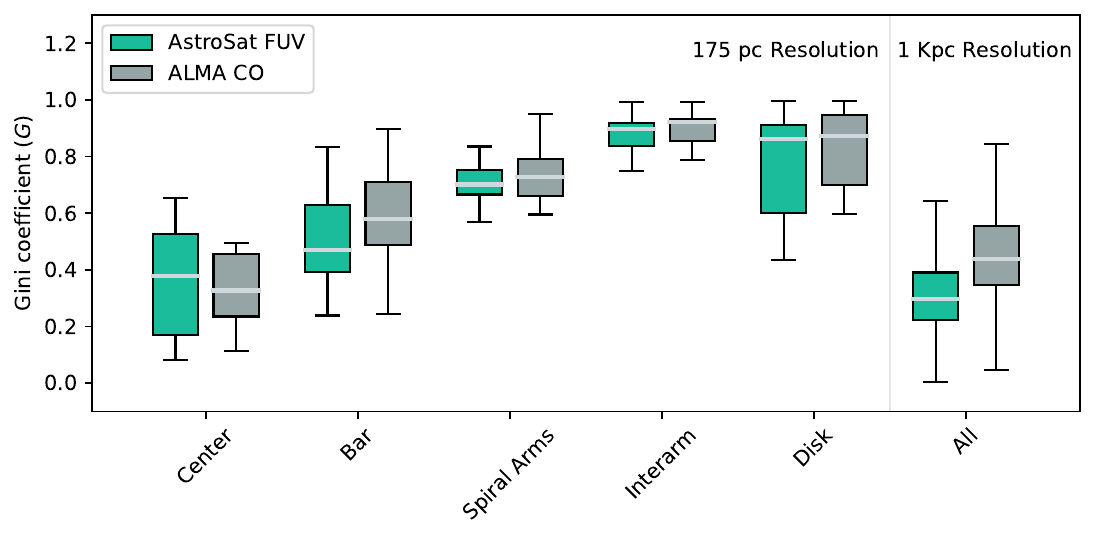}
    \caption{The lower and upper distribution of smoothness measurements of different morphological structures of galaxies at 175\,pc resolution and 1\,kpc size hexagonal apertures, clumping factor (left), and Gini coefficient (right). The horizontal line inside each box plot represents the median value.}
    \label{fig:clumping_env}
\end{figure*}

We use the environment masks from \cite{mask} to measure the clumping factor in different morphological regions of galaxies. The environment masks include almost all morphological features of galaxies, most importantly, spiral arms, interarm regions, and centers. The masks were constructed using visual inspection of IRAC 3.6~$\mu$m data and structural decomposition analysis \citep{bar,salo}. We note that the widths of the spiral arms masks are relatively broad (1-2\,kpc) and are estimated from the spatial distribution of CO emission.

We find that the median values of the clumping factor estimated from FUV and CO emission are similar across most environment masks at a spatial resolution of 175\,pc. This is illustrated in Fig.~\ref{fig:clumping_env}, which shows the range of clumping factors seen in different galaxies that share common morphological features. 

However, the center and spiral arms are the only regions that are more clumped in FUV relative to the CO. The median of the FUV clumping factor is 1.51, whereas the clumping factor for CO is 1.35 in the center of galaxies. To analyze the differences between these two distributions further, we applied the Kolmogorov-Smirnov (K-S) statistics. This analysis resulted in a KS value of 0.4 in the center, indicating a modestly significant difference between the distributions with a p-value of 0.04. In contrast, within the spiral arms, the median clumping factor is $c=2.7$ for the FUV and $c=2.5$ for CO. However, the KS value in this region is lower than in the center, at 0.21. The associated high p-value of 0.92 shows no evidence for a significant difference between the FUV and CO distributions in the spiral arms. Consequently, this suggests that the distribution of the clumping factor for FUV and CO in the spiral arms is more similar compared to the center.

The bars of galaxies have on average the largest clumping factor, particularly for the CO emission, with a CO clumping factor distribution reaching up to $\approx 7$. This result is not surprising since the most massive giant molecular clouds are located in bars \citep[e.g.][]{rosolowsky21} but these regions can also be devoid of molecular gas \citep[][]{Leroy2021}. CO emission in bars has also been shown to have a relatively thin structure \citep{sohpia}.


Using Table~\ref{tab:galpropss}, we observed a higher c$_\mathrm{FUV}$ in the centers of galaxies hosting AGNs. Notably, a similar trend was also seen in a few non-AGN galaxies.


On the other hand, spiral arms are more clumped in FUV emission, implying UV bright complexes are more concentrated and molecular clouds more uniformly spread along spiral arms. The clumping factor could also be exaggerated by high attenuation in the FUV band, broadening the distribution of light. 
Our results suggest that in general, clumps of molecular gas are more concentrated in galaxies, whereas FUV emission is more spread out at 175~pc resolution.

Figure \ref{fig:clumping_env} also shows the distribution of the measured clumping factor using 1~kpc size hexagonal apertures, labeled as ``all''. Similar to the galaxy-integrated values, the median CO clumping factor is still higher ($c=1.7 \pm 0.1$) than the FUV values ($c=1.3 \pm 0.1$). Our results are in good agreement with \cite{Sun_2022}, where they used highly-sensitive ALMA observations of 90 nearby galaxies and reported $c_\mathrm{CO}=1.9$ within hexagonal apertures.  \cite{Leroy_2013} reported a higher CO clumping factor ($c=7$) at 20-300~pc resolution. One should account for molecule-poor low-metallicity systems that \cite{Leroy_2013} included in their sample, which usually have very high clumping factors.  In contrast, the PHANGS-ALMA sample typically studies more distant, higher mass systems (i.e., more metal rich and bright in CO).

\subsection{Gini coefficient}

We also use the Gini coefficient ($G$) as a non-parametric  measure of the distribution of a quantity \citep{gini2}. It has been adapted for quantitative morphological studies of galaxies at different wavelengths, including GALEX observations \cite[e.g.][]{gini_class}. It measures the inequality in a distribution and the higher values indicate a concentration of medium brightness in a small region. The non-normalized values of the clumping factor might raise a concern where two different galactic environments exhibit similar values but not the same morphology. Hence, the Gini coefficient is considered here to assist us in distinguishing between the smoothness of different environments. Following \citet{Davis}, we define the parameter as:

\begin{equation}
G_{\theta\,\mathrm{pc}} = \frac{1}{\bar{I}\, n (n-1)} \sum_{i=1}^{n} (2i-n-1)I_{i},
\label{eq:gini}
\end{equation}
where $n$ is a number of pixels, $I_{i}$ surface brightness of each pixel sorted in ascending brightness order, and $\bar{I}$ is the mean brightness measured over all the pixels. The summation ($i$) is over all pixels of the area. The Gini coefficient, with values between 0 and 1, measures the concentration of clumps in an area. A higher $G$ value indicates a greater concentration of clumps. The accuracy of the $G$ measurement can be influenced by factors such as aperture size and the signal-to-noise ratio of the data. If there are many faint pixels in the region, the results may be impacted \citep{Lisker}. We present the distribution of the Gini coefficient in Fig.~\ref{fig:clumping_env}. The median of the Gini coefficient measured in 1\,kpc hexagonal apertures shows that our measurements of $G_{\rm CO}=0.43$ are higher than $G_{\rm FUV}=0.3$. This aligns with clumping factor measurements, indicating that UV emission is smoother at a kpc resolution.

The Gini coefficient presents a complementary picture of smoothness of the ISM in galaxies compared to the clumping factor. As illustrated in Fig.~\ref{fig:clumping_env}, the smoothness of both the CO and FUV emission decreases in parallel from the inside to the outer parts of galaxies. The centers exhibit a smoother distribution than spiral arms and disks. We note that the high clumpiness measured by Gini in disks are due to the high level of noise in ALMA data at 175~pc.  In spiral arms, there is no clear correlation between $G$ and $c$, with $G$ suggesting a more clumpy distribution. This discrepancy may be due to the sensitivity of $G$ to the effect of faint diffuse emission in spiral arms or the limited sample size (only 14 galaxies with spiral arms masks at this resolution). However, both measures find that the clumping of the CO and FUV emission is similar with the CO being typically slightly more clustered than the FUV.  Such a trend could be expected given the typical ``cycling'' time of the molecular ISM \citep[$<30$~Myr][]{Blitz80,chevance22} is less than the FUV emission that would be produced from the stars formed from the molecular clouds (100~Myr).  Such clumping should be readily measured in simulations.


\section{Comparing FUV and H$\alpha$ Emission}
\label{sec:fuvha}

In this section, we compare the FUV to the H$\alpha$ emission in the resolved structure of galaxies. FUV and H$\alpha$ are tracers of star formation on different timescales. The UV continuum arises primarily from the photosphere of O and B stars ($\mathrm{M}>3~\mathrm{M}_{\odot}$), sensitive to star formation over a 100~Myr time scale \citep{kennicutt12}. In contrast, the H$\alpha$ emission line originates from the recombination of gas that is ionized in \ion{H}{2} regions by massive stars (Extreme UV emission), which are typically O and only early-type B stars with masses of $M>17~M_{\odot}\, $\citep{lee2009, lee11}. The H$\alpha$ emission line traces recent star formation with a timescale of about 5\, Myr \citep{kennicutt12}.  Hence, the H$\alpha$ emission is the optical proxy for photons with $E_\gamma > 13.6~\mathrm{eV}$ but the {\it AstroSat} FUV data trace photons with $6 < E_\gamma/\mathrm{eV} < 9$.

Both tracers also suffer the effects of dust attenuation, but the attenuation toward the shorter wavelengths (FUV) is higher. Many studies \cite[e.g][]{Calzetti94,Kreckel_2013} have observed that nebular emission from dust-enveloped \ion{H}{2} regions undergo roughly double the reddening of the stellar continuum. As a result, H$\alpha$ attenuation is more comparable to the UV continuum, once differential reddening between gas and stars is factored in \citep{lee2009}. 

Utilizing the improved resolution from the {\it AstroSat} FUV observations, our goal is to compare these observations with PHANGS-MUSE H$\alpha$ data \citep{phangs-muse}. With 15 targets common to both data sets, we can probe the variation in the FUV-to-H$\alpha$ ratio across distinct galactic environments at 175\,pc and 1\,kpc resolution. This ratio is influenced by a multitude of factors such as dust attenuation, the age and mass of the stellar populations, star formation history, Initial Mass Function (IMF), and metallicity \citep{lee2009}.

The objective of this section is to examine the local variations of $\log$ FUV/H$\alpha$ (hereafter the ``log ratio'') with respect to the resolved structure of galaxies at a scale of 175~pc.  Using the original unit of the UVIT maps (erg\,cm $^{-2}$\,$\lambda^{-1}$\,s$^{-1}$), in this section, we express our results for the log ratio in terms of:
\begin{align}
    \log\left(\frac{\mathrm{FUV}}{\mathrm{H}\alpha}\right) &= \log_{10} \left(\frac{\lambda_{\mathrm{FUV}} F_\mathrm{FUV}/\mathrm{(erg \, s^{-1}\,cm ^{-2})}}{F_\mathrm{H\alpha} / \mathrm{(erg\,s^{-1}\,cm^{-2})}}\right) \\
    &= \log_{10} \left(\frac{ L_\mathrm{FUV}/\mathrm{(erg\,s^{-1})}}{L_\mathrm{H\alpha} / \mathrm{(erg\,s^{-1})}}\right) 
\end{align}
where the $\lambda_\mathrm{FUV}$ is the central wavelength of the FUV filter used.

The aim is to build upon the previous study by \cite{lee2009} on an integrated galaxy scale and understand which physical properties affect the log ratio on a sub-kpc scale.
 Our analysis centers on the observed log FUV/H$\alpha$ ratio with an emphasis on the effect of attenuation. However, since we do not have NUV observations for all of our targets, we cannot independently assess the stellar attenuation at UV bands. 
 

\begin{figure}[!t]
    \centering
    \includegraphics[width=\columnwidth]{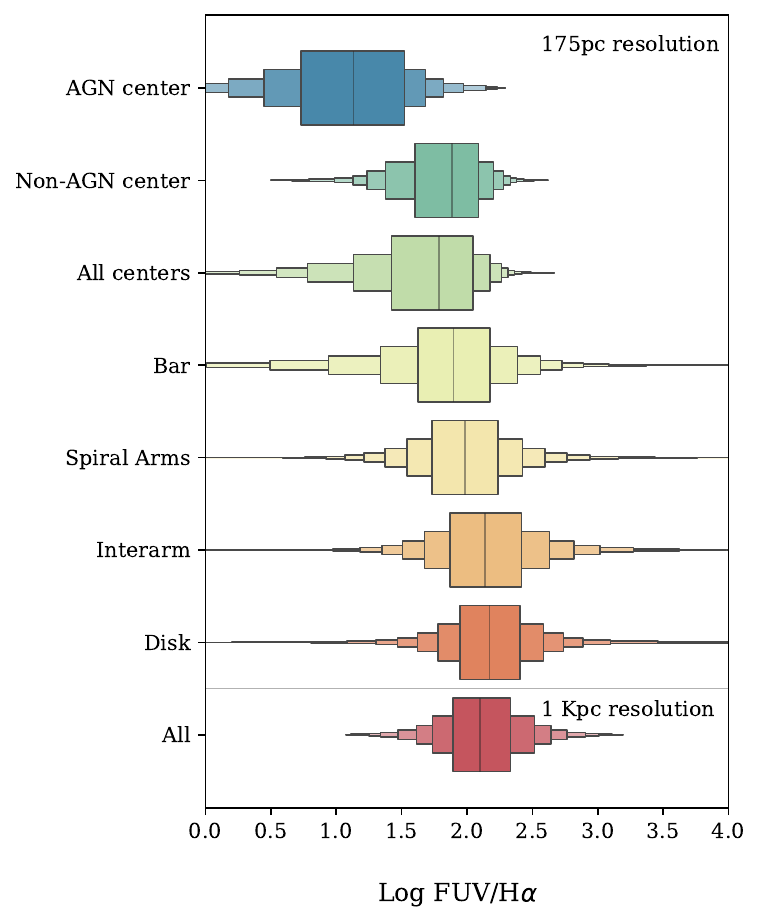}
    \caption{Distribution of the log FUV/H$\alpha$ ratio across different morphological environments using both 175~pc resolution and 1~kpc hexagonal apertures (each box represents different quantiles of the ratio). The tails of the distributions reveal variations in fainter quantiles, particularly a lower fraction in centers and bars. The median of each distribution is marked with a vertical line in the central quantile.}
    \label{fig:fuv_ha_ratio_env}
\end{figure}

\subsection{Environmental dependency}\label{ssec:env}

We investigate the variation of the log FUV/H$\alpha$ ratio across different environments by examining the distribution of the uncorrected log FUV/H$\alpha$ ratio at resolutions of 175~pc and 1~kpc, as shown in Fig.~\ref{fig:fuv_ha_ratio_env}. The median of log FUV/H$\alpha$ ratio is approximately 2 across these environments. At the coarser 1~kpc resolution, the log ratio exhibits a median of 2.1 within hexagonal apertures for all targets. These observations align well with predictions from stellar population synthesis codes. In these models, the H$\alpha$ flux is determined under the assumption of Case B recombination without flux leakage, and FUV fluxes are derived by stacking stellar spectra for each cluster. Notably, \cite{Hermanowicz2013} reported \( \log(\mathrm{FUV/H}\alpha) \) values in the range of 1.5 to 2.5.

The centers of galaxies have a median of $\log(\mathrm{FUV/H}\alpha) = 1.7$, the lowest among all environments, indicating a greater impact of dust attenuation, older stellar populations, or both. Furthermore, we found spiral arms exhibit a higher median $\log(\mathrm{FUV/H}\alpha) = 1.9$ -- 2.0. The distribution of FUV/H$\alpha$ is similar in inter-arm regions and disks with a median of $\sim 2.1$.



AGN-hosting galaxies, which were discussed in Section \ref{sec:clumpingg}, also exhibit a lower log ratio (1.13) compared to non-AGN galaxies (1.88) in the center, although a direct connection between AGN activity and this ratio has yet to be established. Among the sampled galaxies, the center of NGC 1365 displays the lowest log ratio, with $\log(\mathrm{FUV/H}\alpha)<1$. The classification of the central source in this galaxy, as provided by \cite{mask}, refers to an ellipse with a radius of more than 1.5 kpc, thereby encompassing the central star-forming ring. As highlighted by \cite{Whitmore}, this central ring contains several bright, young star-forming regions, which are notably luminous in narrow-band H$\alpha$ HST maps (private communication with Ashley Barnes). This leads to an increased H$\alpha$ emission in the center, with approximately 70 percent of this emission stemming from star-forming regions rather than the AGN. Therefore, the low FUV/H$\alpha$ ratio observed in this region cannot be attributed exclusively to the AGN's influence. Further investigation of the inner 5\arcsec (= 475~pc) of the NGC 1365 central source reveals a smooth non-star-forming disk, reminiscent of early-type galaxies. This region remains stable against gravitational collapse \citep{Schinnerer}. Other galaxies hosting AGNs, such as NGC 1566, NGC 3627, and NGC 7496, exhibit a log FUV/H$\alpha$ ratio exceeding 1.4 in their centers.
We note that the measurement of H$\alpha$ in the center of AGNs can be challenging, as its spectra show broad features and multiple components that are not fitted by the single Gaussian fitting approach used by \cite{phangs-muse}.

\subsection{Radial profiles}
\label{sec:radial}

\begin{figure}[!t]
    \centering
    \includegraphics[width=\columnwidth]{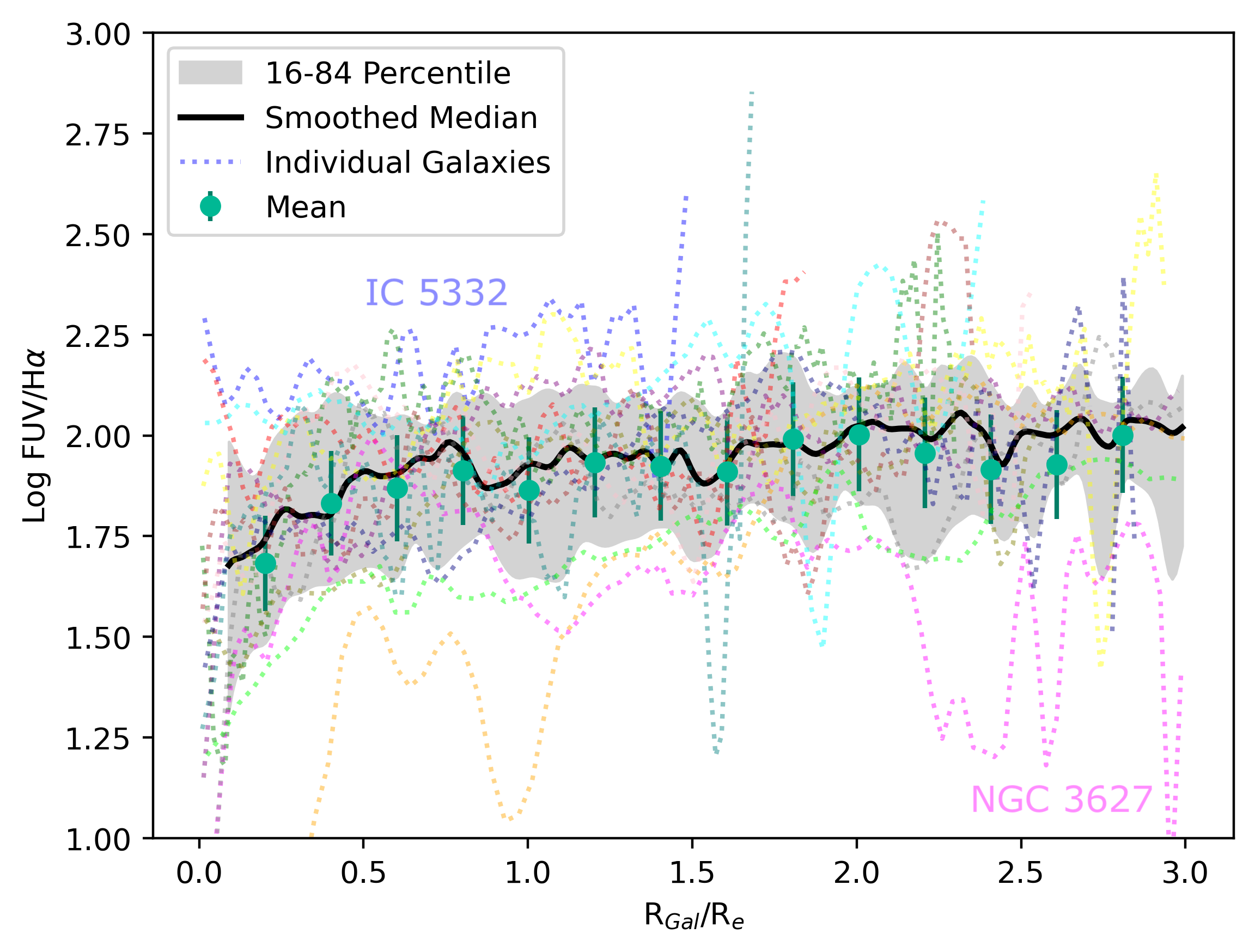}
    \caption{Gaussian-smoothed radial profile of log FUV/H$\alpha$ for all galaxies at a resolution of 175\,pc. The median values are represented by a black line, while the 16-84 percentile range of values is shown in gray. The mean of the smoothed profile is displayed as green circles with error bars, taking into account uncertainties in both H$\alpha$ and FUV. The radial profile of each galaxy is presented with a dashed line. The observations are limited to the field of view of MUSE. The radial profiles of two galaxies that do not exactly follow the median trend are highlighted and discussed in Section \ref{sec:radial}: IC 5332 (purple) and NGC 3627 (pink).}
    \label{fig:fuv_ha_profile}
\end{figure}


Figure \ref{fig:fuv_ha_profile} presents the variation of the log ratio with galactocentric radius scaled to the effective radius ($R_e$) for the 15 galaxies in the sample, obtained from \cite{phangsalma}. The median profile across all 15 targets is nearly flat with a typical value of $\log(\mathrm{FUV/H}\alpha)=1.9$, which is consistent with the analysis we provide in Section \ref{ssec:env}. Most of our targets have a flat radial profile, with a few spikes or dips. Most notably, the galaxies IC5332 (log FUV/H$\alpha \sim 2.1$) and NGC3627 (log FUV/H$\alpha \sim 1.65$), show a significant variation.

The low-mass spiral galaxy IC5332, which is a candidate for XUV type 2 as discussed in Section \ref{sec:xuv}, shows a higher log FUV/H$\alpha$ ratio. The majority of its \ion{H}{2} regions have a ratio between 1.6 and 1.9, with a few regions exhibiting higher ratios ($>$2) in spiral arm ends, indicating the presence of older populations. The higher ratio of log (FUV/H$\alpha$) is linked to the low dust content of IC5332, where most of its \ion{H}{2} regions display minimal reddening, $E(B-V)=0.05-0.1$.

Conversely, NGC3627, which extends up to 3 R$_{\text{Gal}}$/R$_{\text{eff}}$, demonstrates several dips in its radial profile, corresponding to various \ion{H}{2} regions combined with a few Milky Way stars in its spiral arm. This combination results in a much lower ratio. The galaxy hosts a massive star-forming region located at the end of the bar, which shows a higher reddening (E(B-V)$>$1) and molecular gas (I$_{\text{CO}}$$>$200~K~km~s$^{-1}$). In the central region, we observe a ratio of 0.7, which, once again, is associated with high reddening and a substantial amount of molecular gas. The \ion{H}{2} regions display ratios of 0.9-1.1, often correlated with nearby bright molecular clouds with intensities of $I_{\text{CO}}$$>$25~K~km~s$^{-1}$.

The nearly flat radial profile of the logarithmic ratio aligns well with the findings of \cite{Mehta}. In their study, they investigated the dust-corrected FUV/H$\alpha$ ratio in intermediate-mass galaxies (with stellar masses ranging from $10^{9.5}$ to $10^{10.2}~M_{\odot}$) at a redshift of approximately 1. They observed a similar trend persisting up to a normalized galactic radius of $R_{\text{gal}}$/$R_{\text{e}}<1.6$.

\subsection{Physical parameters affecting log FUV/H$\alpha$}
Given the relatively high physical resolution of the PHANGS-{\it AstroSat} data set and the richness of the supporting data, we have a new opportunity to understand what physical factors are driving changes in the log FUV/H$\alpha$ ratios on 175~pc scales. These effects have been explored in previous works based on the integrated properties of galaxies or at coarser resolution \citep{lee2009,Meurer}. While $\log (\mathrm{FUV/H}\alpha)$ could provide good insight into galactic evolution through, e.g., tracing the star formation history \citep{Lomaeva}, the effects of dust attenuation and reddening will alter this ratio, which requires carefully understanding the effects of dust.  There are a myriad of possible factors shaping the FUV/H$\alpha$, so we assess whether these effects are detectable in our sample.  In addition to those explored below, previous works have proposed several possible physical factors that change FUV/H$\alpha$ including the porosity of the interstellar medium (ISM), stochastic effects from the number of massive stars, and variations in the Initial Mass Function. Below, we focus our work on factors where we have good observational proxies.

\paragraph{Metallicity} Stellar populations with lower metallicity have hotter atmospheres, resulting in the production of more UV photons, particularly ionizing ones \citep{lee2009}. Conversely, metal-rich populations, such as those concentrated in galaxy centers, tend to be cooler and more evolved, emitting a reduced amount of ionizing flux compared to non-ionizing UV continuum. Theoretical demonstrations by \cite{lee2009}, employing models from \cite{Bruzual}, showed that the effect of metallicity primarily influences H$\alpha$ emission rather than FUV emission due to assumptions regarding case B recombination and the spatial treatment with ionizing photons. However, to observe this effect, one requires a significant variation in metallicity, so we expect this effect to be weak across our sample.

\paragraph{The Equivalent Width (EW) of H$\alpha$} The EW of H$\alpha$ is also used to trace the age of young stellar populations, particularly in areas where the H$\alpha$ emission is bright enough \citep{fabian}. The EW of H$\alpha$ serves as an indicator of the relative brightness of the nebular emission compared to the underlying stellar continuum. Younger stars, with ages on the order of millions of years, exhibit high EW values for H$\alpha$ as they are actively ionizing the surrounding gas, producing intense nebular emission. Observations in both the local universe and at higher redshifts confirm changes in the EW of H$\alpha$ with the age of \ion{H}{2} regions \citep[e.g.,][]{ew,ew_z}. On the other hand, evolutionary synthesis models predicted its functional form \citep{ew2}. The EW of H$\alpha$ is a highly observable signature of the balance between very young and older stars. At the scale of \ion{H}{2} regions, we expect to observe an anti-correlation between the EW of H$\alpha$ and $\log (\mathrm{FUV/H}\alpha)$, where bright H$\alpha$ is associated with ages below 10 Myrs. For older stellar populations, where the EW of H$\alpha$ is below 10 \AA, the relation is not notable.

\paragraph{Star Formation History} Sharp fluctuations in the star formation rate (SFR) could also result in changes in the log FUV/H$\alpha$ ratio, given the varying lifetimes of O and B stars. Consequently, this ratio may be used to trace variations in the star formation rate on different timescales, ranging from 5 to 200~Myr \citep{kennicutt12}. This is often referred to as the ``burstiness'' of star formation. On the other hand, a temporary decrease or ``gasp'' in the star formation rate should also be considered in this context \citep{Meurer}. \citet{Mehta} investigated the time-evolution of the dust-corrected log FUV/H$\alpha$ and found under conditions of steady, constant star formation, an equilibrium state is achieved with a log FUV/H$\alpha \sim 2$. This highlights how SFH parameters such stellar mass, age, and metallicity might impact the FUV/H$\alpha$ ratio in galaxies.
\\


\begin{figure*}
    \centering
\includegraphics[width=0.95\textwidth]{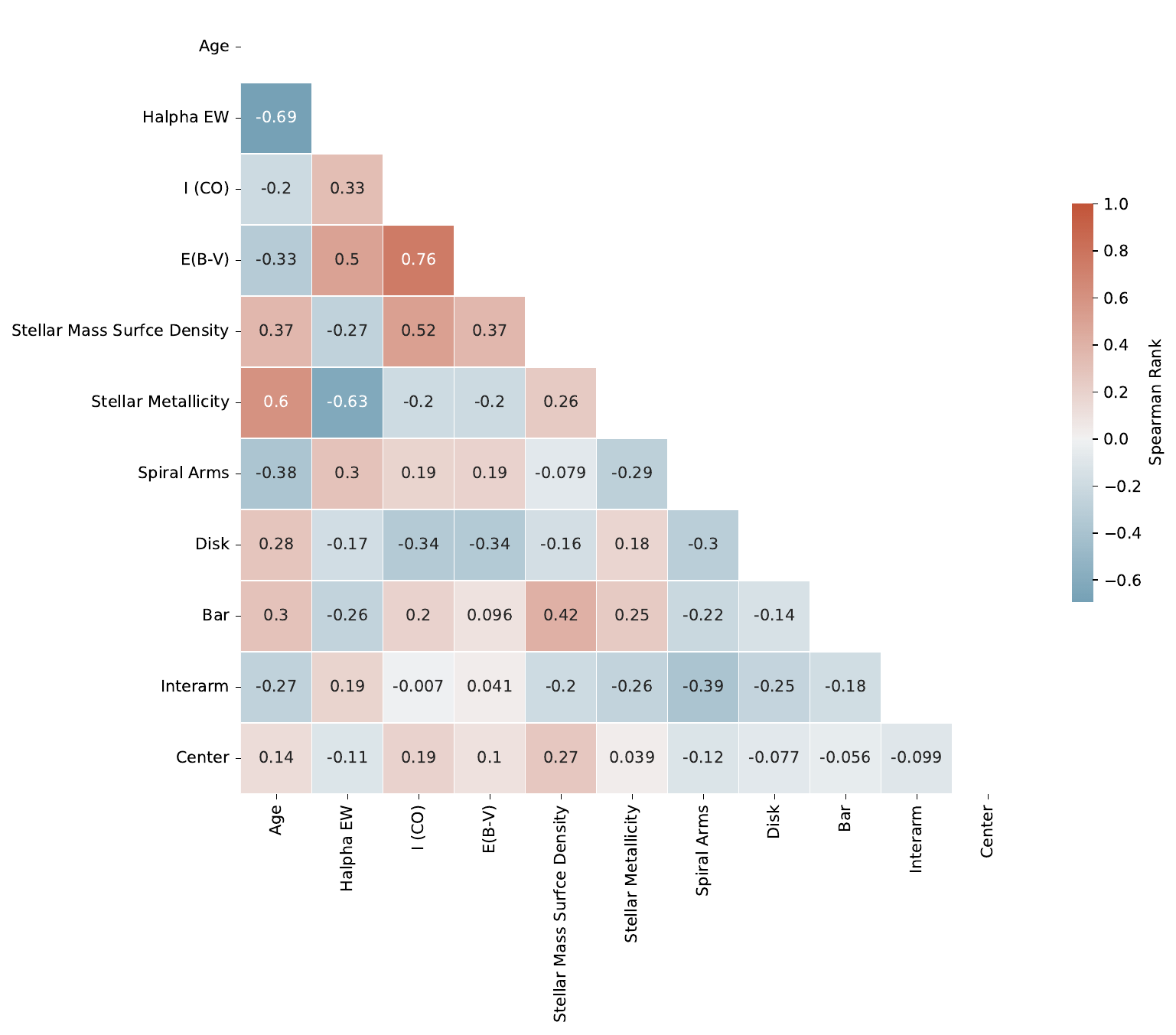}
   \caption{Spearman's correlation matrix illustrating the interrelationships between our predictor variables, analyzed at a resolution of 175~pc.}
    \label{fig:corr}
\end{figure*}

Here we aim to quantify the relationship between $\log(\mathrm{FUV/H}\alpha)$ and various observed quantities. Initially, we introduce a comprehensive model encompassing the majority of observed and physical quantities to describe the log ratio as outlined in Section \ref{subsubsection:lasso}. Next, we present a simplified version of the model wherein dust reddening emerges as the primary factor accounting for 50 percent of the variation in the log ratio. 



\subsubsection{Drivers of the FUV/H$\alpha$ Ratio}
\label{subsubsection:lasso}

In this section, we present a comprehensive model to quantify the role of different variables in predicting the log of the FUV/H$\alpha$ ratio. Previous studies have demonstrated how various physical parameters of galaxies, specifically resolved structural parameters, can be modeled using multivariate linear regression in logarithmic space \cite[e.g.,][]{day,Ssun_2022}. 
We aim to predict log FUV/H$\alpha$ using variables such as E$(B-V)$, $I_\mathrm{CO}$, H$\alpha$ EW, stellar age, mass, and metallicity using the panchromatic PHANGS data. Leveraging the PHANGS MUSE observations, we employ maps of stellar mass, light-weighted age, and light-weighed stellar metallicity for our targets. The specific models used in that fit set the range of expectations for our results, which we summarize here.  Specifically, they employ a \textit{young} fitting template over observed lines and continuum described in \cite{phangs-muse}. In that study, the E-MILES stellar population models from \cite{Vazdekis} and \citet{asad} are expanded to include younger age groups. This is particularly important as we are making a direct comparison of the ratio of young stellar populations traced by H$\alpha$ at $<5$~Myr, to intermediate ages traced by FUV emission. The selected templates incorporate the Padova isochrones and include 18 bins for ages ranging from 6.3 Myr to 14.12~Gyr. Here, we note that most of the ages measured from PHANGS-MUSE using the \textit{young} template are older than 10 Myrs, which does not align with the typical age of 5 Myrs for H$\alpha$ bright \ion{H}{2} regions as highlighted by \citet{kennicutt12,Hassani_2022}. Therefore, to more accurately trace the ages of these younger regions, we have incorporated the EW of H$\alpha$ emission as a weighting factor in the LASSO analysis. Additionally, the stellar metallicity is binned from $-0.7$ to 0.22 for older stellar populations (age $>63$~Myr), while for younger populations, the [Z/H] bins range from $-0.7$ to 0.41. Bright stars are masked in our results to avoid potential distortions.  For our 175-pc resolution, we also include an environmental classification, which can assume values of 0 or 1. These environments are categorized as center, bar, spiral, interarm, and disk. We perform the same analysis for the 1-kpc scale but exclude the environmental classifications since the different environments are not well resolved at this resolution.

We note two important cautions here. Firstly, we assume a linear relationship in log space between each predictor and the outcome variable, although we acknowledge that may not be the case. The combination of dust reddening with stellar mass, age, and metallicity, may not exhibit a linear relation with the log FUV/H$\alpha$, especially at a 1~kpc scale. The interplay and degeneracy between dust, metallicity, and age of stars has been addressed in previous studies \citep{lee2009}. Moreover, due to this degeneracy, we anticipate colinearity between parameters, particularly between $E(B-V)$ and CO intensity, or age and metallicity. This co-linearity will impact our results, as the coefficients of these parameters in the model will be influenced by each other. 

Figure \ref{fig:corr} highlights the correlation between different predictors. We observe a correlation between stellar metallicity and age, with a Spearman coefficient of approximately 0.6, indicating an older population with higher metal content. We also find a strong anti-correlation between the EW of H$\alpha$ and the age of the stellar population. A similar anti-correlation is also observed between the EW of H$\alpha$ emission and stellar metallicity. It is important to note that different environments do not exhibit any significant correlation with physical or observed parameters. However, we do observe a weak trend between the presence of a bar and the surface density of stellar mass, which can be attributed to the concentration of stellar gas in the bar of most galaxies \citep{sohpia}.


We use the Least Absolute Shrinkage and Selection Operator regression \citep[LASSO;][]{lasso_paper} to understand how each predictor control the variation of the log ratio. Following \cite{day}, we use 5-fold cross-validation (CV) to fine-tune the hyperparameters of the model, ensuring that the performance is evaluated on different subsets of the data. LASSO is a regularization method that helps prevent overfitting in linear regression models by introducing a penalty to the absolute sizes of the coefficients. This not only helps in model simplification but also in feature selection, as some coefficients can be shrunk to zero, effectively excluding that predictor from the model. The degree of regularization is controlled by the parameter $\alpha$. By adjusting the $\alpha$ value through cross-validation, we make sure it provides the best fit between our training and validation data. 
This approach decreases the chances of overfitting and enhancing simplifcation of the model. We apply LASSO regression to our linear model, which is represented as:
\begin{equation}
\label{eq2}
\log \left(\frac{\mathrm{FUV}}{\mathrm{H}\alpha}\right) = {\alpha_0} + \sum_{i=1}^{n} \alpha_i [\log X_i - \log\langle X_i \rangle].
\end{equation}
Here, $\alpha_{0}$ represents the intercept, and $\alpha_{i}$ denotes the power-law exponents of our predictors. We standardize our predictors by normalizing by the median values of each variable ($\langle X_i\rangle$). As indicated by Equation \ref{eq2}, we use a log transform for all parameters, but we do not apply the transform for E(B-V) and metallicity, which are already intrinsically logarithmic.
 
\begin{deluxetable*}{llllll}
\tablecaption{
The LASSO regression coefficients ($\alpha_{i}$) for different predictors of log FUV/H$\alpha$ at 175\,pc and 1\,kpc resolution. The norm is the median of each predictor which we subtract that in the log scale before regression. \label{tab:model2}}
\tablehead{
\colhead{Parameters} & \multicolumn{2}{c}{175 pc} & \multicolumn{2}{c}{1 kpc} \\
& \colhead{Norm} & \colhead{$\alpha_{i}$} & \colhead{Norm} & \colhead{$\alpha_{i}$}  
}
\startdata
$E(B-V)$  &  0.17~$\mathrm{mag}$ & $-0.86 \pm 0.01 $ & 0.11 $\mathrm{mag}$ & $-0.61 \pm 0.1$ \\
$I_\mathrm{CO}$ & 2.51~$\mathrm{K~km~s^{-1}}$ & $-0.11 \pm 0.01$& 1.75~$\mathrm{K~km~s^{-1}}$ & $-0.10 \pm 0.02$\\
EW(H$\alpha$) &15~\AA & $-0.36 \pm 0.01$ &11~\AA & $-0.46 \pm 0.02$ \\
Age  &  615~Myr & $-0.30 \pm 0.01$& 707~Myr & $-0.36 \pm 0.03$ \\
$\Sigma_\star$ & 174~$M_\odot~\mathrm{pc}^{-2}$ & $-0.07 \pm 0.01 $ & 140~$M_\odot~\mathrm{pc}^{-2}$ & $-0.18 \pm 0.04$ \\
$[\mathrm{Z/H}]$ & $-0.29$ dex & $0.06\pm0.01$ & $-0.22$~dex & 0.00 \\
\hline
Environments\tablenotemark{a}  & & \\
\hline
Center &\nodata & $-0.03 \pm 0.01 $ & \nodata  & \nodata \\
Bar & \nodata & $-0.08 \pm 0.01$ & \nodata  & \nodata \\
Spiral & \nodata & 0.00 & \nodata  & \nodata \\
Interarm &\nodata & $-0.02 \pm 0.01$ & \nodata  & \nodata \\
Disk & \nodata & $-0.02 \pm 0.01 $ & \nodata  & \nodata \\
\hline
$R^2$ & \multicolumn{2}{c}{0.67} & \multicolumn{2}{c}{0.66}  \\
 Residual (dex) & \multicolumn{2}{c}{0.18} &  \multicolumn{2}{c}{0.20}
\enddata
\tablenotetext{a}{Using environmental masks, we identify different morphological regions of galaxies and assign a weight of 0 or 1 to each individual part.\\}
\tablecomments{The residual value represents the scatter around the model, where we use the standard deviation to measure it. The VIF for all parameters ranging from 2 to 3.7. Notably, the VIF for stellar mass is 5.8 at 1 kpc scale.}
\end{deluxetable*}
We present results of the regression at 175 kpc and 1 kpc resolution in Table \ref{tab:model2}. The intercept term ($\alpha_0$) represents the expected log FUV/H$\alpha$ when all predictors are equal to their medians, which is $\approx2.0$ at both resolutions. The slope coefficients indicate the change in the log FUV/H$\alpha$ while holding all other predictors constant. At 175~pc resolution, more localized effects and variations within resolved structures of galaxies may introduce additional complexities, which could account for the observed differences. In general, we observe negative exponents for all predictors, except for metallicity. A higher coefficient for dust reddening is observed at both resolutions, indicating the importance of attenuation in the log ratio.  We also note that the slopes for the different environment predictors are all weak, suggesting that variation in the log ratio that we observe can be explained better by variations in other parameters in the model.  Similarly, the scaling for metallicity ($Z/H$) is also weak, though the range of metallcities probed is relatively small.


Additionally, the H$\alpha$ EW also exhibits a negative slope, suggesting that younger sources with prominent H$\alpha$ EW values may result in a lower log FUV/H$\alpha$ ratio. However, it is important to note that the relationship between the age of the stellar population and the H$\alpha$ EW is not linear \citep{ew}. We note the intrinsic correlation between the line brightness and its EW, which is tight in star-forming regions.

Furthermore, we observe a negative slope in the relationship between age and the FUV/H$\alpha$ ratio, which is surprising. On the scale of \ion{H}{2} regions or stellar clusters (i.e., $<50$~pc), we expect that a higher FUV/H$\alpha$ ratio would correspond to an older population (i.e., $>100$~Myr) where the H$\alpha$ emission is faint \citep{Ujjwal}. However, at a resolution of 175~pc, which is the best common resolution available, there might not be a clear relation due to the mixing of diffuse H$\alpha$ emission with \ion{H}{2} regions. Taking into account the degeneracy between age, metallicity, and dust attenuation, we propose that this negative slope may result from non-linear relations between other parameters correlated with the stellar age.
To clarify how this could have originated from LASSO interpretation, we propose a simplified model to evaluate its performance in the next section.

To assess the uncertainty associated with the estimated coefficients in the regression results reported in Table \ref{tab:model2}, we employ a bootstrap resampling method. This approach involves generating multiple bootstrap samples by repeating the same LASSO regression with cross-validation. We calculate the standard errors of the slopes by computing the standard deviation and across the iterations. Our findings reveal that the standard deviation on $\alpha_i$ at a scale of 1~kpc is below 0.1 for all predictors. Similarly, the standard deviations are below 0.01 at 175~pc scale. 

\subsubsection{Simplified model}

The LASSO regression develops a multifactor model and the method does not isolate the drivers of the log(FUV/H$\alpha$) ratio automatically.  However, it does show some factors have a stronger influence on the variation (large $\alpha_i$) than others.  Here, we develop a simple model that explains most of the variation in the FUV/H$\alpha$ ratio.

\begin{figure}[!t]
    \centering
    \includegraphics[width=\columnwidth]{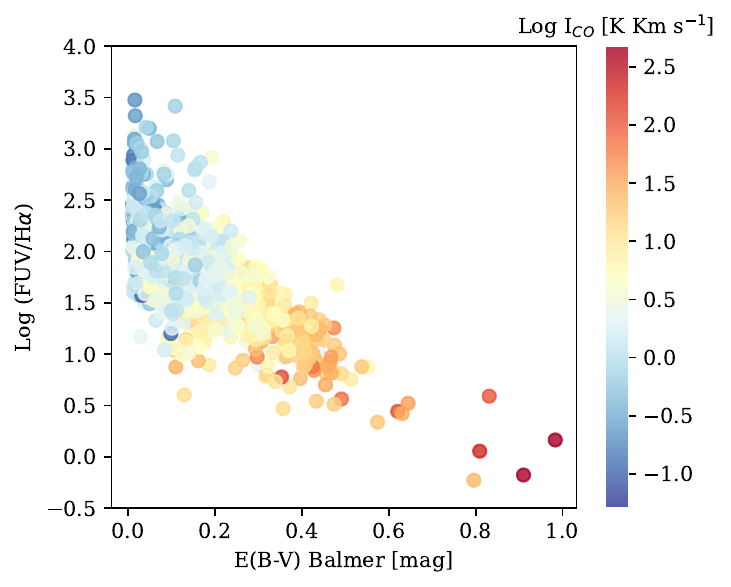}
    \caption{Log of FUV/H$\alpha$ versus the Balmer decrement E(B-V) in 1~kpc size hexagonal apertures. The reddening is measured using MUSE/VLT observations in different galaxies. The colorbar represents the CO(2-1) line integrated intensity. The figure shows that $\log(\mathrm{FUV/H\alpha})$ depends on both $E(B-V)$ and $I_\mathrm{CO}$.}
    \label{fig:fuv_ha_ratio_dust}
\end{figure}

The strongest factor in the analysis is dust attenuation as measured by the color excess $E$(B-V), suggesting the observed variations in FUV/H$\alpha$ could be driven by attenuation alone. To delve deeper and find a simplified model to describe the log ratio, we run a model that includes only dust reddening, E(B-V). This resulted in dust reddening coefficients of approximately $-1.7$ at 175 pc scale. The goodness of fit, represented by $R^{2}$, yielded values of 0.5. This indicates that dust reddening alone can account for roughly half of the variation in the log ratio. Notably, a standalone model of CO can predict the log ratio as effectively as a single reddening component, with an $R^{2}$ of 0.4. However, the coefficient of the CO component is $-0.4$. Figure \ref{fig:fuv_ha_ratio_dust} illustrates that a lower $\log(\mathrm{FUV/H}\alpha)$ ratio (indicating higher attenuation) corresponds to higher CO intensity, particularly in regions containing dense, dusty, and massive star-forming regions. This trend holds not only on the 1~kpc scale, as shown in Figure \ref{fig:fuv_ha_ratio_dust}, but also at the 175~pc resolution. This suggests that dust and molecular emission are tracing each other.  This supposition is borne out through measuring a significant correlation between CO emission and $E(B-V)$ on a kpc scale, supported by a Spearman rank coefficient of $\rho \approx 0.75$. This correlation remains robust even when we enhance the resolution to 175 pc, although it exhibits notable variations across diverse environments. 

\begin{deluxetable*}{llll}
\tablecaption{
Regression coefficients for simplified models predicting $\log(\mathrm{FUV/H\alpha})$.\label{tab:model3}}
\label{tab:simpleres}
\tablehead{
\colhead{Parameters} & Normalization & \colhead{$\alpha_{i}$}  & \colhead{$R^2$}
}

\startdata
Observed FUV/H$\alpha$ (175pc scale) \\
\hline
$E(B-V)$  & 0.17$~\mathrm{mag}$ & $-1.68\pm0.01$ & \multirow{2}{*}{0.50} \\
$\alpha_0$ & 1.95 & \\
\hline
$I_\mathrm{CO}$ & 2.51$~\mathrm{K~km~s^{-1}}$ & $-0.41 \pm 0.01$ & \multirow{2}{*}{0.41} \\
$\alpha_0$ & 1.94 & \\
\hline
$E(B-V)$ & 0.17$~\mathrm{mag}$ & $-1.22\pm0.01$ &   \multirow{3}{*}{0.51} \\
$I_\mathrm{CO}$ &2.51$~\mathrm{K~km~s^{-1}}$ & $-0.15 \pm 0.01$ &   \\
$\alpha_0$ & 1.95 & \\
\hline 
$E(B-V)$ & 0.17~$\mathrm{mag}$ & $-1.38\pm0.01$  &   \multirow{2}{*}{0.56}  \\
EW(H$\alpha$)  & 15~\AA& $-0.17\pm0.01$ & \\
$\alpha_0$  &  1.94 & \\
\hline 
$E(B-V)$ &0.17~$\mathrm{mag}$&  $-1.38\pm0.01$ & \multirow{4}{*}{0.64}  \\
EW(H$\alpha$) & 15~\AA & $-0.34\pm0.01$& \\
Age  & 615~Myr &$-0.34\pm0.01$ & \\
$\alpha_0$ & 1.93 & \\
\hline
Attenuation-corrected FUV/H$\alpha$ (1 kpc scale) \\
Age  & 840$~\mathrm{Myr}$ & $0.84\pm0.02$ & \multirow{3}{*}{0.49} \\
$\Sigma_{*}$  & 168$\, M_\odot~\mathrm{pc}^{-2}$ & $-1.39\pm0.02$  \\
$[Z/H]$  & -0.19  dex & $0.33\pm0.02$  \\
$\alpha_0$ & 2.98 & \\
\enddata
\end{deluxetable*}

Furthermore, we find that the combination of reddening and H$\alpha$ EW led to a better prediction of the model, increasing $R^{2}$ to 0.56. Interestingly, a model that includes only reddening and ages behaves very similarly, with the same $R^{2}$ as the previous  model, indicating that H$\alpha$ EW effectively traces the age of the stellar population. We note that even in the simplified models, the coefficient for age remains negative. Finally, including both age and H$\alpha$ EW in the model increased both $R^{2}$ values to over 0.6, making them very similar to our comprehensive model. This also suggests that the inclusion of stellar mass, metallicity and different environmental factors has a negligible effect on controlling the log ratio.

Based on the proposed simplified model of a single dust reddening component, we have successfully predicted the FUV/H$\alpha$ ratio with an $R^{2} \sim 0.5 $ at scales of 1~kpc and 175~pc, respectively. Subsequently, the simplified model exhibits higher scatter, approximately 0.2 to 0.25 dex, due to the inclusion of fewer parameters. However, the collinearity among various parameters complicates the interpretation of the individual effects of each parameter on the log ratio. We summarized the simplified models and their coefficients in Table \ref{tab:model3}.

\subsubsection{Dust attenuation correction to FUV/H$\alpha$}
\label{subsec:dust_correction}

\begin{figure}[!t]
    \centering
    \includegraphics[width=\columnwidth]{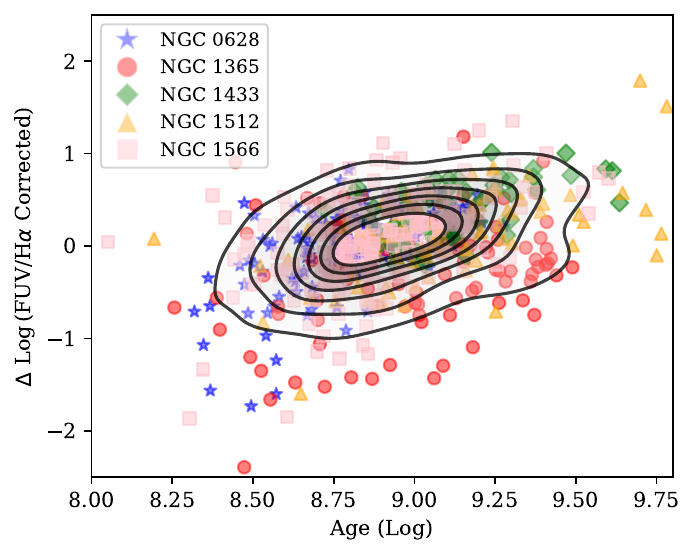}
    \caption{The residual values of log FUV/H$\alpha$ versus age, corrected for internal attenuation and adjusted by removing the influences of stellar mass and metallicity. The black contours show a Kernel Density Estimation of the observed data dataset, focusing on the data distribution between the 16th to 84th percentiles. This relation includes measurements from only the five galaxies with NUV data available, which are necessary to measure the slope of the UV continuum and hence the FUV attenuation correction. }
    \label{fig:kde_plot}
\end{figure}

As demonstrated in the previous section, dust attenuation is the primary factor governing variations in the uncorrected log FUV/H$\alpha$. In this section, we attempt to correct  FUV/H$\alpha$ for dust attenuation and explore which parameter(s) exert the most influence on this log ratio.

Using the available NUV AstroSat observations for five of our targets, we first measure the slope of the UV continuum ($\beta$) to correct the FUV emission for internal attenuation. Following \cite{Boquien_2012}, we define $\beta$ as follows:
\begin{equation}
    \beta = \frac{\text{log} ({F}_\text{FUV} / {F}_\text{NUV})}{\text{log} (\lambda_\text{FUV}/ \lambda_\text{NUV})} -2 
\end{equation}
with $F$ representing the flux density and $\lambda$ the central wavelength of the UVIT filter. We use a prescription following \cite{Boquien_2012} to estimate $A_{FUV}$ using our measured $\beta$ values, thereby correcting the FUV emission.  We also use the Balmer decrement measurements and $E(B-V)$ presented previously to correct the H$\alpha$ emission, based on a typical value of the total-to-selective value $R_{V}=3.1$ \citep{Calzetti2000,Belfiore}. After correcting both FUV and H$\alpha$, we measure the log ratio using 1 kpc hexagonal apertures and measure how well the variation of  ratio is predicted by the luminosity-weighted  age,  surface mass density, and metallicity of the stellar population. We found that this model can describe the variation of the dust-attenuation corrected log ratio with an $R^{2} = 0.49$. Our results show coefficients of 0.84 for age, $-1.39$ for the stellar mass surface density, and 0.33 for the stellar metallicity, along with an intercept of $\alpha_0 = 2.98$ (c.f.\ Table \ref{tab:model3}). Similar to the approach used in \citet{day}, we perform a partial regression and define the residual of this parameter ($\Delta \log \mathrm{FUV/H}\alpha$ corrected), as:

\begin{equation}
\begin{aligned}
    \Delta\, \log\, (\text{FUV}/\text{H}\alpha)_\text{corr} &= \log\, (\text{FUV}/H\alpha)_\text{corr} - \alpha_{0} \\
    &\quad - (-1.39 \log \Sigma_{*} - \langle \Sigma_{*} \rangle) \\
    &\quad - (0.33 [Z/H] - \langle [Z/H] \rangle)
\end{aligned}
\end{equation}
where $\langle \Sigma_{*} \rangle$ and $\langle [Z/H] \rangle$ are median normalization factors, and $\alpha_{0}$ is the intercept reported in Table \ref{tab:simpleres} for the dust-attenuation corrected log ratio. This relation effectively removes the influence of stellar mass surface density and metallicity from the dust-attenuated log ratio and highlights its variation with age. Figure \ref{fig:kde_plot} shows the positive correlation between ($\Delta \log \mathrm{FUV/H}\alpha$ corrected) and stellar age across five galaxies. FUV emission can come from less massive stars with longer
lifespans, unlike the more massive, short-lived stars (typically less than 10 million years) responsible for H$\alpha$ emission \citep{kennicutt12}. Hence, the ratio should increase with the average age of a
galaxy's stellar population, which in turn depends on the galaxy's star formation history.

The star formation histories, whether constant or bursty, will further influence the evolution of the log ratio. After a burst of star formation, the FUV/H$\alpha$ ratio will quickly rise, due to the sudden drop in H$\alpha$ emission. Thereafter, the ratio will decrease again, moving towards an equilibrium value as the FUV-emitting stars eventually die.  The models of \citet{Mehta} show that the log of the FUV/Ha ratio will reach an equilibrium state $100-200$ million years after the onset of a constant star formation history (see also the model predictions from \cite{Weisz} for log M$_{\star}\sim10$ galaxies).  The ages shown in Figure \ref{fig:kde_plot} shows the luminosity-weighted measurement of ages from \cite{phangs-muse}, which trace the integrated star formation histories and show longer absolute timescales (log(age)$\sim9$).

\section{Conclusion}
\label{sec:conc}
We present the {\it AstroSat} UVIT observations of 31 massive spiral nearby galaxies spanning over an order of magnitude in stellar mass. These galaxies are part of the PHANGS Survey of nearby galaxies which is synthesizing a broad range of multiwavelength data, including the ALMA and VLT/MUSE data that support this work.  UVIT data improves over previous UV observations with $\sim 1.4''$ resolution enabling studies of young stellar populations and dust attenuation on  $<200$~pc scales. This atlas contains data available from dedicated observations collected as part of PHANGS as well as a uniform reduction of the available data from the AstroSat archive.  Hence, the atlas contains data from galaxies in one to nine UV bands covering wavelengths from 1480~\AA\, to 2790~\AA. These data can provide new resolved observational constraints to challenging questions such as attenuation curve, dust attenuation, and UV SFR for 30\% of the full PHANGS sample. The PHANGS-AstroSAT atlas is available online\footnote{\url{https://www.canfar.net/storage/vault/list/phangs/RELEASES/PHANGS-AstroSat/v1p0}}. This paper describes:
\begin{enumerate}
  \item \textit{Observations and quality assessment:}  The PHANGS-{\it AstroSat} atlas targets low-inclination ($i<75^\circ$) massive ($M_\star>10^{9.57}~M_\odot$, actively star-forming galaxies in the local universe (D$<22\,$Mpc). As described in Section \ref{sec:data}, we observe 11 galaxies for $\sim 3$ ks each in the F148W band.  For these and 20 other targets available in the archive, we apply the standard UVIT data calibration processes in \textsc{CCDLAB} \citep[Section \ref{sec:procs};][]{postma17}. We found that the pipeline reduction yields a typical resolution of 1.4\arcsec\ at the FUV and 1.2\arcsec\ at the NUV band. We provide science-ready maps of our targets after background subtraction and foreground extinction corrections.  The UV background emission is $\sim10^{-19}$~erg~s$^{-1}$~cm$^{-2}\, $\AA$^{-1}$ for most of our targets at $\lambda =1480$\,\AA. We assess the quality of our data by comparing \textit{AstroSat} UVIT observations with GALEX maps from \cite{z0mgs}. The comparison of our data set with GALEX observations shows good flux scale agreement ($<10\%$ variations), as illustrated in Figs.~ \ref{fig:fuv_galex},~\ref{fig:nuv_galex}.
 
  \item \textit{XUV disks}: In Section \ref{sec:xuv} we present NGC 6744 as a new Type 1 XUV disk candidate and IC~5332 as a  candidate for a Type 2 XUV disk galaxy.  NGC~6744 shows a very large scale length in FUV and we identify two bright spiral arms beyond 1.5 $R_{25}$, without any R-band or H$\alpha$ emission.  We did not find convincing evidence other candidates that were previously unknown though we confirm all previous classifications \citep{Thilker2007}.
  

  \item \textit{Structure of CO and FUV emission}: In Section \ref{sec:clumping}, we quantify the clumping and smoothness of the FUV and CO emission using two non-parametric scalar measurements: the clumping factor ($c$) and the Gini coefficient ($G$).  Using more than 4000 1-kpc hexagonal apertures in our targets, we find $c_\mathrm{CO}=1.7$ versus a median of c$_\mathrm{FUV}=1.3$ , illustrating the FUV emission is less clumped than the CO.  In general, we find that bars are the most clumped environments. Although the median clumping factor from FUV and CO emission is similar across different environments, CO clumping factors are slightly higher than FUV except in the center and spiral arms. Furthermore, clumps of UV bright emission, which could have \ion{H}{2} region counterparts, have the most effect on increasing the clumping factor. The non-parametric Gini coefficient also predicts a smoother distribution for FUV emission compared to CO except in the center, which is in agreement with clumping factor results. 

  \item \textit{FUV to H$\alpha$ ratio}: In pairing UVIT and MUSE H$\alpha$ maps, we used 1\,kpc size apertures and see a $\sim$2~dex difference in non-dust-corrected FUV and H$\alpha$ luminosity (Section \ref{sec:fuvha}). We find $\log_{10}(\mathrm{FUV/H}\alpha)<2$ the centers and bars. The log ratio at the center of AGN galaxies is 65 percent lower than that in normal star-forming galaxies. The radial profile of $\log_{10}(\mathrm{FUV/H}\alpha)$ is approximately flat outside centers, and shows small variation from 1.7 to 2.1.  We propose an empirical model to predict the observed log ratio in our sample.  The model considers how several parameters predict FUV/H$\alpha$ including CO molecular gas emission,  $E(B-V)$, star formation history parameters (mass, age, metallicity), and the equivalent width of H$\alpha$.  A full model is able to explain the variation of $\log_{10}(\mathrm{FUV/H}\alpha)$ at resolutions of 175~pc and 1~kpc, achieving approximately $R^{2}\sim0.65$. In evaluating different versions of the model, we argue that dust attenuation is the dominant factor controlling the variation of the log ratio in star forming disk galaxies. Finally, having corrected both FUV and H$\alpha$ emissions for internal dust attenuation, we find that when the influences of stellar mass and metallicity are excluded from the FUV/H$\alpha$ log ratio, a positive correlation emerges between this ratio and the age of the stellar population.
\end{enumerate}
The new perspective on star-forming complexes offered by {\it AstroSat} UVIT, when combined with our other $\sim 1\arcsec$ resolution surveys such as PHANGS-ALMA and PHANGS-MUSE, provides fresh insights on UV emission in nearby galaxies. Future work will be able to investigate hybrid UV+IR star formation rate calibrations, the attenuation curve, and the relation of the excess of dust to UV luminosity versus the stellar color (IRX-$\beta$) at sub-kpc parsec scales over a broad population of nearby galaxies.

\software{python, Astropy \citep{astropy,astropy2,astropy2022}, numpy \citep{numpy}, CCDLAB \citep{postma17}, spectral-cube \citep{spectral-cube}, and matplotlib \citep{matplotlib}, ChatGPT.}

\begin{figure*}[t!]
    \centering
    \includegraphics[width=0.7\textwidth]{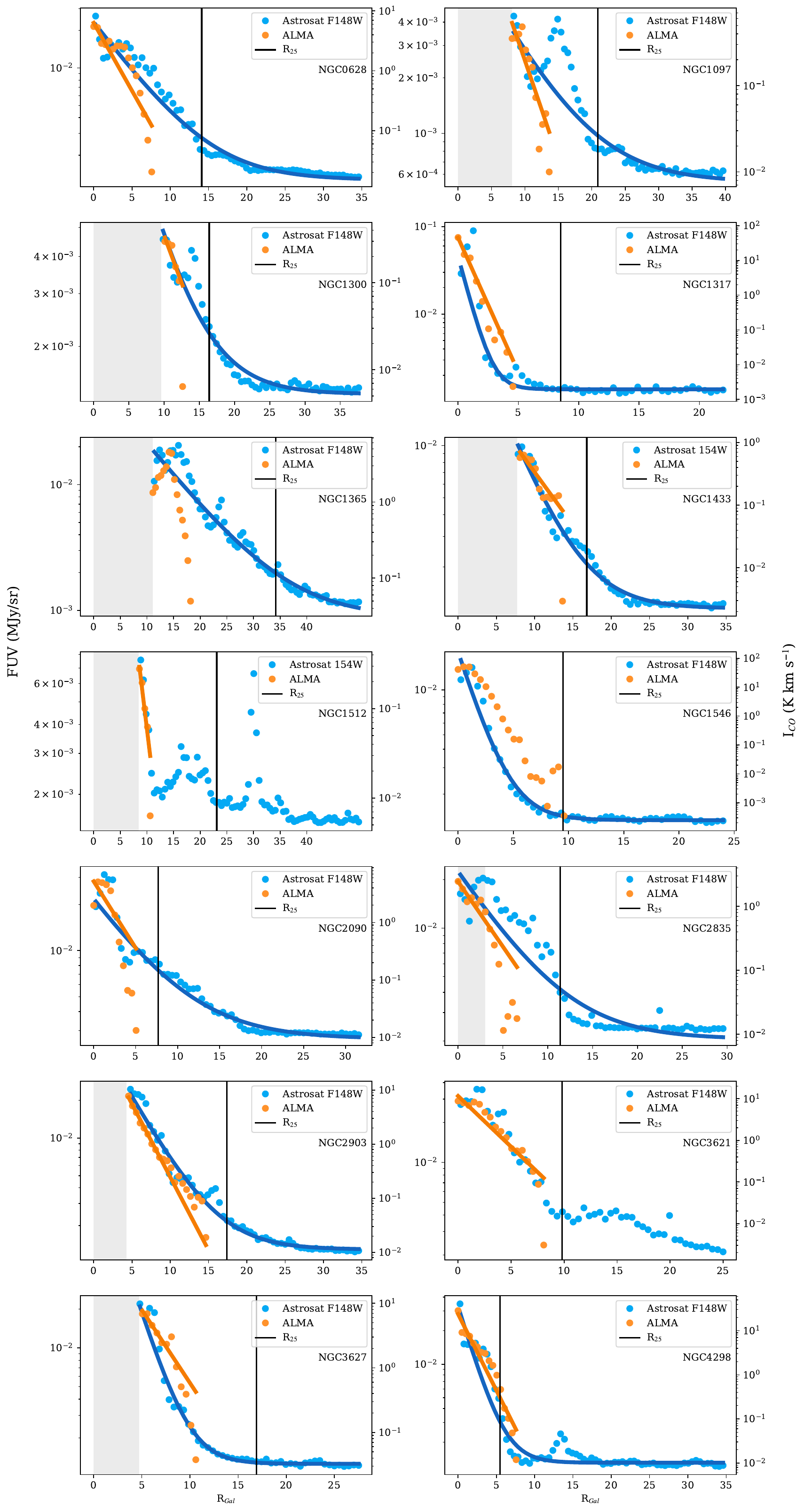}
  \caption{Same as the Fig \ref{fig:radial_profile1}.}
    \label{fig:radial_profile2}
\end{figure*}

\begin{figure*}[t!]
    \centering
\label{fig:ap}
  \includegraphics[width=0.74\textwidth]{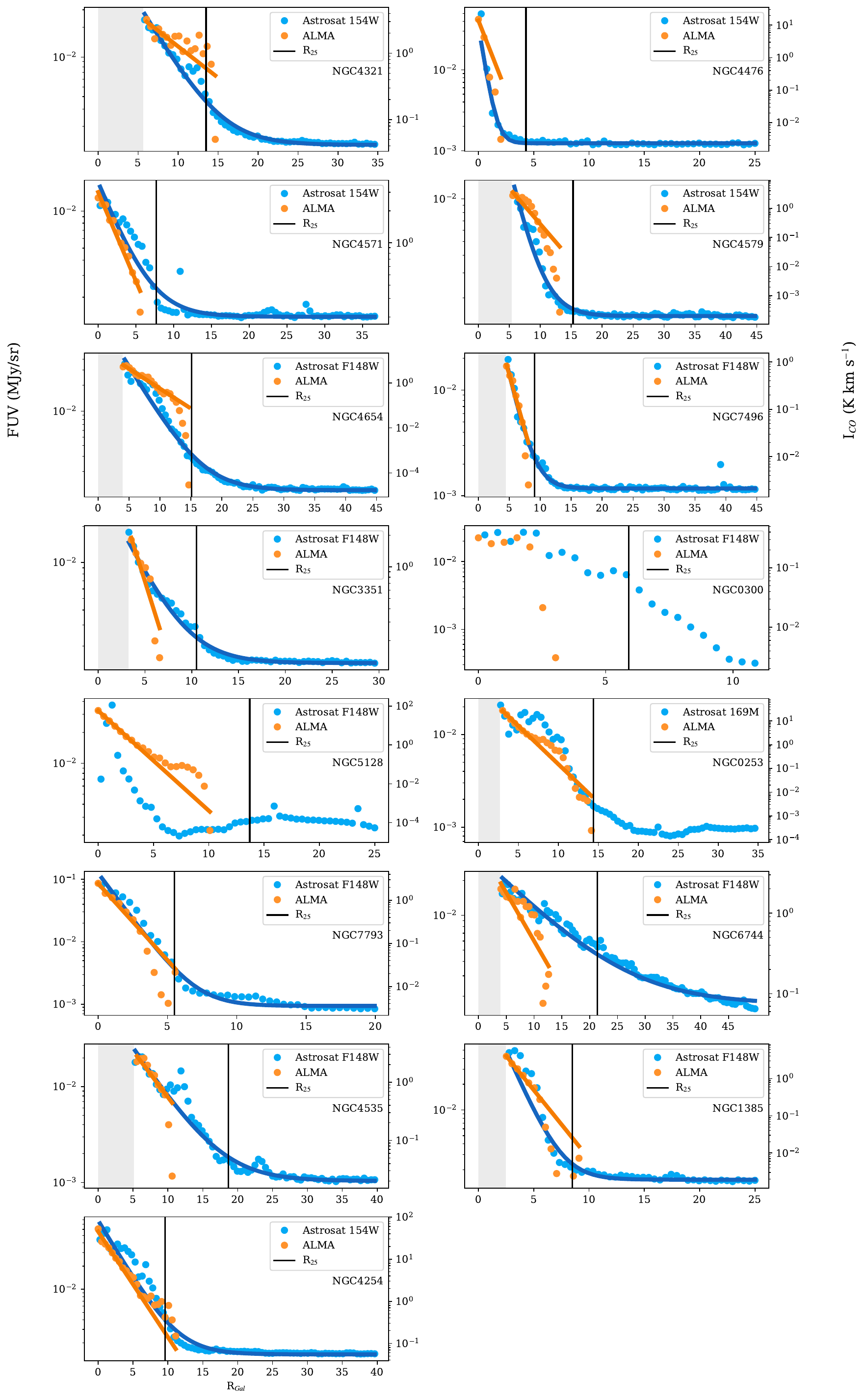}
  \caption{Same as the Fig \ref{fig:radial_profile1}.}
    \label{fig:radial_profile3}
\end{figure*}

\appendix
\section{\\ Radial profiles}
\label{ap1}
Figures \ref{fig:radial_profile2},\ref{fig:radial_profile3} present the radial profiles for all PHANGS targets that are not presented in Fig. \ref{fig:radial_profile1}.

\begin{acknowledgments}

We are grateful for the comments of an anonymous referee that improved the quality of this paper. This work was conducted as part of the PHANGS collaboration. This publication uses the data from the {\it AstroSat} mission of the Indian Space Research Organisation (ISRO), archived at the Indian Space Science Data Centre (ISSDC). HH, ER, JN, EWK, and HC acknowledge the support of the Natural Sciences and Engineering Research Council of Canada (NSERC), funding reference number RGPIN-2017-03987, and the Canadian Space Agency funding reference numbers SE-ASTROSAT19, 22ASTALBER.
EWK acknowledges support from the Smithsonian Institution as a Submillimeter Array (SMA) Fellow and the Natural Sciences and Engineering Research Council of Canada.
RSK and SCOG acknowledge funding from the European Research Council via the ERC Synergy Grant ``ECOGAL'' (project ID 855130), from the German Excellence Strategy via the Heidelberg Cluster of Excellence (EXC 2181 - 390900948) ``STRUCTURES'', and from the German Ministry for Economic Affairs and Climate Action in project ``MAINN'' (funding ID 50OO2206). The team in Heidelberg also thanks for computing resources provided by {\em The L\"{a}nd} and DFG through grant INST 35/1134-1 FUGG and for data storage at SDS@hd through grant INST 35/1314-1 FUGG. MB gratefully acknowledges support from the ANID BASAL project FB210003 and from the FONDECYT regular grant 1211000. JMDK gratefully acknowledges funding from the DFG through an Emmy Noether Research Group (grant number KR4801/1-1), as well as from the European Research Council (ERC) under the European Union's Horizon 2020 research and innovation programme via the ERC Starting Grant MUSTANG (grant agreement number 714907). 
COOL Research DAO is a Decentralized Autonomous Organization supporting research in astrophysics aimed at uncovering our cosmic origins.
MC gratefully acknowledges funding from the DFG through an Emmy Noether Research Group (grant number CH2137/1-1). HAP acknowledges support by the National Science and Technology Council of Taiwan under grant 110-2112-M-032-020-MY3.
KK, OE gratefully acknowledge funding from the Deutsche Forschungsgemeinschaft (DFG, German Research Foundation) in the form of an Emmy Noether Research Group (grant number KR4598/2-1, PI Kreckel) and the European Research Council’s starting grant ERC StG-101077573 (“ISM-METALS"). 
KG is supported by the Australian Research Council through the Discovery Early Career Researcher Award (DECRA) Fellowship (project number DE220100766) funded by the Australian Government. 
KG is supported by the Australian Research Council Centre of Excellence for All Sky Astrophysics in 3 Dimensions (ASTRO~3D), through project number CE170100013. 
J.K. is supported by a Kavli Fellowship at the Kavli Institute for Particle Astrophysics and Cosmology (KIPAC).
HH acknowledges the use of ChatGPT to refine the English-language presentation of the ideas in this paper.

\end{acknowledgments}

\bibliographystyle{aasjournal}
\bibliography{main}



\end{document}